\documentclass[11pt,a4paper]{article}
\pdfoutput=1 
\usepackage[margin=1in,includefoot]{geometry}
\usepackage{amsmath,amssymb}
\usepackage[mathscr]{euscript}

\usepackage[pdfstartview=FitH,pdfpagemode=None]{hyperref}
\usepackage{cite}
\usepackage{braket}

\usepackage{authblk}

\usepackage{xcolor}

\title{Vacuum configuration of winding superstrings\\from non-standard semiclassical quantization}
            
\author[a,b]{Tommaso Canneti}
\author[c,d]{Federico Castellani}
\author[e,f]{Wolfgang M\"uck}
 
\affil[a]{Dipartimento di Fisica, Universit\`a di Torino \protect\\ Via Pietro Giuria 1; 10125 Torino, Italy}
\affil[b]{Istituto Nazionale di Fisica Nucleare, Sezione di Torino \protect\\ Via Pietro Giuria 1; 10125 Torino, Italy}
\affil[c]{Dipartimento di Fisica e Astronomia, Universit\`a di Firenze \protect\\ Via G. Sansone 1; 50019 Sesto Fiorentino (Firenze), Italy}
\affil[d]{Istituto Nazionale di Fisica Nucleare, Sezione di Firenze \protect\\ Via G. Sansone 1; 50019 Sesto Fiorentino (Firenze), Italy}
\affil[e]{Dipartimento di Fisica ``Ettore Pancini'', Universit\`a degli Studi di Napoli Federico II\protect\\ Via Cintia; 80126 Napoli, Italy}
\affil[f]{Istituto Nazionale di Fisica Nucleare, Sezione di Napoli \protect\\ Via Cintia; 80126 Napoli, Italy}
\date{\small{tommaso.canneti@unito.it, federico.castellani@unifi.it, mueck@na.infn.it}}

\numberwithin{equation}{section}

\newcommand{\ie}{i.e.,\ }
\newcommand{\eg}{e.g.,\ }

\DeclareMathOperator{\sgn}{sgn}

\newcommand{\rmd}{\mathrm{d}}
\newcommand{\rmD}{\mathrm{D}}

\newcommand{\Tr}{\operatorname{Tr}}

\newcommand{\e}[1]{\operatorname{e}^{#1}}

\newcommand{\hide}[1]{}

\newcommand{\unih}{\mathfrak{h}}
\newcommand{\unie}{\mathfrak{e}}
\newcommand{\unig}{\mathfrak{g}}

\newcommand{\vev}[1]{\left\langle #1 \right\rangle}

\newcommand{\pull}[1]{\lfloor #1}

\newcommand{\geleven}{\Gamma^{(11)}}
\newcommand{\gtwo}{\Gamma^{(3)}}
\newcommand{\gnine}{\Gamma^{(9)}}

\begin{document}
\maketitle

\begin{abstract}
{\small \noindent Semiclassical quantization techniques are unreliable, if the magnitude of the quantum fluctuations becomes comparable with a classical background scale. This is the case, for example, for superstrings at temperatures close to the Hagedorn temperature. Here, we propose a non-standard treatment which aims to extend the semiclassical approach for type-II superstrings to such cases. This allows us to provide a well-posed reformulation of recent computations of the Hagedorn temperature in a large class of confining gauge theories. Along the way, we fill a gap in the literature by proving that the standard semiclassical partition function of the Green-Schwarz superstring can be made finite, by means of a suitable choice of integration measures, for any background worldsheet in a target spacetime satisfying the (type-IIA) supergravity field equations.
}
\end{abstract}

\newpage
\tableofcontents

\linespread{1.3}

\section{Introduction}

The semiclassical quantization of strings and branes is an established method to test the AdS/CFT correspondence \cite{Maldacena:1997re, Gubser:1998bc, Witten:1998qj} beyond the leading order. Integrability \cite{Beisert:2010jr, Bombardelli:2016rwb}, the superconformal bootstrap \cite{Belavin:1984vu, Beem:2013qxa}, and supersymmetric localization \cite{Witten:1988xj, Pestun:2007rz} provide exact results on the CFT side, which can be compared to string theory computations order by order in large-$N$ and large coupling expansions, for example, of Wilson loops \cite{Drukker:2000ep, Drukker:2011za, Forini:2012rh, Aguilera-Damia:2014bqa, Bergamin:2015vxa, Forini:2015mca, Forini:2015bgo, Faraggi:2016ekd, Cagnazzo:2017sny, Forini:2017whz, Aguilera-Damia:2018bam, Aguilera-Damia:2018rjb, Aguilera-Damia:2018twq, David:2019lhr, Medina-Rincon:2019bcc, Gautason:2021vfc, Giombi:2020mhz, Astesiano:2024sgi}.  
Certain non-supersymmetric and non-conformal gauge theories \cite{Witten:1998zw} also possess holographic duals, opening the possibility to study more realistic scenarios from the gravity side. It is thus clear that the semiclassical quantization of strings represents an important tool and deserves to be studied in all its subtleties.

The object of our study is the partition function of the Green-Schwarz superstring\footnote{Strictly speaking, the partition functions should involve also the sum over worldsheet topologies, which we ignore by considering the topology as fixed.}
\begin{equation}
	\label{intro:path.int} 
	Z = \int \frac{\mathcal{D} h \mathcal{D}X \mathcal{D}\theta}{\mathcal{V}ol} \e{-S_P[h,X,\theta]}~,
\end{equation} 
where $S_P[h,X,\theta]$ is the Green-Schwarz type-II(A or B) superstring action functional, written in the Polyakov formalism with an auxiliary metric $h^{\alpha\beta}$ \cite{Cvetic:1999zs, Tseytlin:1996hs}, 
\begin{equation}
	\label{intro:GS.action}
	S_P = -\frac1{4\pi \alpha'} \int \rmd^2 \sigma \left( \sqrt{-h} h^{\alpha\beta} G_{\alpha\beta} + \varepsilon^{\alpha\beta} B_{\alpha\beta} \right)~.
\end{equation}
Here, $G_{\alpha\beta}$ and $B_{\alpha\beta}$ are the string worldsheet metric and the Kalb-Ramond 2-form, respectively, which depend on the embedding of the string worldsheet in the target superspace. The two contributions to the action are respectively called the metric and Wess-Zumino terms. $\varepsilon^{\alpha\beta}$ is the 2-d Levi-Civita symbol (not the tensor). In \eqref{intro:path.int} the functional integrations are carried out over the superspace coordinates of the embedding, $X(\sigma)$ and $\theta(\sigma)$, as well as over the auxiliary metric, modulo symmetries, which are worldsheet diffeomorphisms, Weyl invariance, and the fermionic kappa symmetry. This is formally indicated by the factor $\mathcal{V}ol$ in the denominator. We will not be more specific about the integration measures knowingly ignoring possible subtleties that may arise along the way. 

Our aim is to study \eqref{intro:path.int} in a semiclassical approximation quantizing the string fluctuations around a classical  worldsheet embedded in an arbitrary on-shell 10-d supergravity field configuration. As there already is a consistent body of literature on this topic, starting from papers from the 1980s \cite{Polyakov:1981rd, Polyakov:1981re, Polchinski:1985zf, Polyakov87, Kavalov:1985di, Fradkin:1985ys, Fradkin:1985fq, Wiegmann:1989md} and including the many more recent papers on Wilson loops mentioned earlier, let us first explain why we feel that there is space for revisiting this topic. We essentially have two reasons. First, to the best of our knowledge, the procedure of semiclassical quantization of superstrings has always been developed for 10-d supergravity backgrounds, in which some fields have been assumed to be absent, mostly because they really were absent in the specific cases of interest. The semiclassical expansion of the bosonic part of the superstring action featuring a non-trivial Kalb-Ramond field, but without the fermionic sector, which couples to the Ramond-Ramond fields, has been carried out in \cite{Singh:2023olv}. 
We will, therefore, consider the general case including all supergravity background fields.
We shall present the type-IIA case, the type-IIB case can be obtained with appropriate substitutions \cite{Wulff:2013kga}. To explain our second and main motivation, we point out that, in a standard semiclassical calculation, it is irrelevant whether one uses the Polyakov or the Nambu-Goto formulations, the latter formally being obtained from the former by integrating out the auxiliary metric in \eqref{intro:path.int}. This equivalence has been spelled out in very nice detail in \cite{Drukker:2000ep, Forini:2015mca}, including a discussion of diverse gauge choices. Now, it is clear that any semiclassical calculation is ``valid'' as long as the typical size of the quantum fluctuations is parametrically small compared to any classical background (length) scale. Recently, however, the semiclassical method has been pushed to its limit in the context of a string worldsheet approach to the Hagedorn temperature in confining gauge theories \cite{Bigazzi:2022gal, Bigazzi:2023oqm}, where a certain classical scale parameter (the compactification radius) is naturally small and comparable in size with the quantum fluctuations. Not surprisingly, it was found that the next-to-leading (NLO) term in the Hagedorn temperature (in an expansion in terms of the inverse 't~Hooft coupling) calculated in the standard semiclassical approach was off by a certain factor when compared to an integrability-based calculation. However, it was also noted in \cite{Bigazzi:2023oqm} that the correct NLO term could be obtained by a change of the fluctuations' mass parameter, which formally would correspond to keeping the auxiliary metric off-shell. The role of the fermions was assumed just to assure the cancellation of the divergent terms in the one-loop effective action \cite{Bigazzi:2024biz}, just as it happens in the standard setup. Treating a background field as classically off-shell is, of course, not the standard semiclassical approach. In fact, the calculation in \cite{Bigazzi:2024biz} leaves open some technical issues, which we aim to address with our work.

Let us emphasize that our aim is not to quantize the entire superstring in a 10-d supergravity geometry, which may be achieved in certain fortunate cases by coset constructions, see \cite{Demulder:2023bux} and references therein. Our target is just the semiclassical effective action for a given classical string worldsheet, which is much more modest.

The outline of the paper is as follows. In section~\ref{NambuGoto}, the semiclassical expansion of the type-IIA superstring in the Nambu-Goto formulation will be considered in all generality, \ie for any 10-d background that satisfies the IIA supergravity field equations. After introducing the dynamics of the physical fluctuations, which consist of eight scalars parameterizing the normal fluctuations of the worldsheet and eight 2-d spinors, we study the structure of the generic logarithmic divergence of the semiclassical effective action. It will be shown how this divergence can be cancelled by suitable choices of the integration measures.
In section \ref{Polyakov}, we turn to the Polyakov formulation and first establish its equivalence to the Nambu-Goto formulation in the standard semiclassical setting. Then, we introduce the non-standard approach, which consists in keeping the worldsheet auxiliary metric classically off-shell. This change of scheme has profound and generally problematic implications, which we will explain. Special attention is paid to the problems arising in connection with fixing the kappa symmetry in this setting. 
As an application of the non-standard semiclassical approach, we consider in section~\ref{Hage} a class of superstrings located at special loci in what are known as ``cigar'' geometries, which are the gravity duals of certain confining gauge theories. We will obtain the ground state configuration of the auxiliary metric, which turns out to be different from the background induced metric, and use this result to identify the Hagedorn temperature of the string and, correspondingly, the confining gauge theory. A number of specific examples are discussed in section~\ref{Examples}.

Finally, section \ref{concl} contains our conclusions, while supporting material can be found in the appendices.

\section{Superstring fluctuations in the Nambu-Goto formulation} 
\label{NambuGoto}

In this section, we will consider the semiclassical fluctuations of superstrings in the Nambu-Goto formulation. For the most part, this is well known material (see, \eg \cite{Drukker:2000ep, Forini:2015mca}, a treatment of the bosonic sector with a background Kalb-Ramond field can be found in \cite{Singh:2023olv}).  Our extension with respect to these and other earlier works is that we allow for arbitrary on-shell (type-IIA) supergravity backgrounds including the dilaton, Kalb-Ramond and Ramond-Ramond fields, for which we prove the finiteness of the effective action with appropriate integration measures. We hope that this exposition may also introduce the reader to the notations we will adopt throughout the paper.

\subsection{String action and background field equation}

The passage from the Polyakov formulation to the Nambu-Goto formulation is achieved by eliminating the auxiliary metric in \eqref{intro:GS.action} via its field equation, which can be expressed as
\begin{equation}
	\label{NG:h.eom}
	h^{\alpha\gamma} G_{\gamma\beta} - \frac12 \delta^\alpha_\beta h^{\gamma\delta} G_{\gamma\delta} =0~,
\end{equation}
so that \eqref{intro:GS.action} becomes 
\begin{equation}
\label{NG:action.NG}
	S_{NG} = -\frac1{2\pi \alpha'} \int \rmd^2 \sigma \left( \sqrt{-G} + \frac12 \varepsilon^{\alpha\beta} B_{\alpha\beta} \right)~. 
\end{equation}
The semiclassical action is obtained by expanding the fields $G_{\alpha\beta}$ and $B_{\alpha\beta}$ in terms of fluctuations around a classical string worldsheet and then truncating the action at second order in the fluctuations. Manifestly covariant expressions are obtained, if this expansion is based on a geodesic expansion (exponential map) or, equivalently, using Riemann normal coordinates \cite{Alvarez-Gaume:1981exa, Forini:2015mca}. 
This procedure, and its analogue in superspace \cite{Atick:1986jr}, emphasizes the geometric nature of the fields. 
It has been described in detail elsewhere \cite{Forini:2015mca, Bigazzi:2023oqm, Wulff:2013kga}, and we refrain from repeating it here. We include in Appendix~\ref{embed} a summary of the geometry of embedded manifolds in a form that naturally includes spinor fields. 

Expanding first to second order in the fermions, $G_{\alpha\beta}$ and $B_{\alpha\beta}$ are given by \cite{Grisaru:1985fv, Wulff:2013kga}
\begin{align}
\label{NG:G}
	G_{\alpha\beta} &= g_{\alpha\beta} + i \bar{\theta} \Gamma_{(\alpha} \mathcal{D}_{\beta)} \theta~,\\
\label{NG:B}
	B_{\alpha\beta} &= b_{\alpha\beta} + i \bar{\theta} \geleven \Gamma_{[\alpha} \mathcal{D}_{\beta]} \theta~,
\end{align}
where $g_{\alpha\beta}$ and $b_{\alpha\beta}$ are the purely bosonic components, and $\mathcal{D}_{\alpha}$ denotes the pull-back onto the worldsheet of the generalized covariant derivative operator 
\begin{equation}
\label{NG:D}
	\mathcal{D}_\mu = \rmD_\mu +\frac18 H_{\mu m n} \Gamma^{mn} \geleven +\frac18 S \Gamma_\mu~,
\end{equation}
with 
\begin{equation}
\label{NG:S}
	S = \e{\phi} \left( \frac12 F_{mn} \Gamma^{mn} \geleven +\frac1{4!} F_{mnpq} \Gamma^{mnpq} \right)~. 
\end{equation}
The analogous expressions valid in the type-IIB case can be found in \cite{Wulff:2013kga}. $\theta$ is a real 32-component (10-d Majorana) spinor.\footnote{To be more precise, we have absorbed the appropriate spinor rotation $U$ that accompanies the rotation of the spacetime frame to a worldsheet adapted frame, see appendix~\ref{embed} for details, into $\theta$. Alternatively, using the local $SO(D)$ freedom, one may always work with a spacetime frame that is adapted to the background worldsheet and simply set $u_m{}^n =\delta_m^n$ and $U=1$.}

The expansion of the bosonic embedding coordinates should be done in form of an exponential map,
\begin{equation}
\label{NG:X.expand}
	X^\mu \to [\exp_X(Y)]^\mu \equiv X^\mu + Y^\mu -\frac12 \Gamma^\mu{}_{\nu\rho} Y^\nu Y^\rho + \cdots~,
\end{equation}
where $X^\mu$ on the right hand side is the background embedding, and the fluctuations $Y^\mu$ form a spacetime vector living on the worldsheet. It decomposes into longitudinal and normal components, 
\begin{equation}
\label{NG:Y.decomp}
	Y^\mu = \zeta^\alpha x_\alpha^\mu + \chi^i N_i^\mu~,
\end{equation}
with a 2-d vector $\zeta^\alpha$ and eight scalars $\chi^i$, which live in the normal bundle and are gauged with respect to local $SO(8)$ rotations. For more details we refer again to \cite{Forini:2015mca, Bigazzi:2023oqm}. 
The 2-d vector $\zeta^\alpha$ is associated with first-order worldsheet diffeomorphisms and most conveniently gauge fixed to zero. 
We call this the \emph{static gauge}, although often this terminology refers to a non-covariant identification of worldsheet and spacetime coordinates. In the Nambu-Goto formulation, this worldsheet covariant static gauge is the natural gauge choice. We remark that it is not really necessary to fix the gauge. In fact, up to quadratic order, the terms in the action that involve the longitudinal fluctuations disappear by virtue of the background field equations or reduce to boundary terms. At higher orders, the action can always be expressed in terms of gauge invariant combinations of the fluctuation fields, which coincide with the normal fluctuations in static gauge.

In static gauge and up to second order in the fluctuations, the action \eqref{NG:action.NG} is
\begin{equation}
\label{NG:action.expand}
	S_{NG} = -\frac1{2\pi\alpha'} \int  \rmd^2 \sigma\, \sqrt{-g} \left(\mathcal{L}_0 + \mathcal{L}_1 + \mathcal{L}_2 +\cdots \right)~, 
\end{equation}
with the Lagrangian densities
\begin{align}
\label{NG:L0}
	\mathcal{L}_0 &= 1+\frac12 \epsilon^{\alpha\beta} b_{\alpha\beta}~,\\
\label{NG:L1}
	\mathcal{L}_1 &=  \left( -K_{i\alpha}{}^\alpha + \frac12 \epsilon^{\alpha\beta}H_{\alpha\beta i} \right) \chi^i~,\\
\label{NG:L2}
	\mathcal{L}_2 &= \frac12 (\nabla^\alpha \chi^i)(\nabla_\alpha \chi_i) 
	-\frac12 \left( R^\alpha{}_{i\alpha j} +K_{i \alpha\beta} K_j{}^{\alpha\beta} - K_{i \alpha}{}^\alpha K_{j\beta}{}^{\beta} \right) \chi^i \chi^j \\
\notag &\quad 
	+ \frac12 \epsilon^{\alpha\beta} \left[ \frac12 x_\alpha^\mu x_\beta^\nu N_i^\rho N_j^\lambda \left(\nabla_\rho H_{\lambda\mu\nu} \right)   
	\chi^i \chi^j + K_{i\alpha}{}^\gamma H_{\beta\gamma j} \chi^i \chi^j +H_{\alpha ij} \chi^j \nabla_\beta \chi^i \right] \\
\notag &\quad	 
	+ \frac{i}2 \bar{\theta} \left( \Gamma^\alpha \mathcal{D}_\alpha + \epsilon^{\alpha\beta} \geleven\Gamma_\alpha \mathcal{D}_\beta \right) \theta~.
\end{align}
All quantities except the scalar fields $\chi^i$ and the spinor $\theta$ just depend on the geometry of the background worldsheet. These quantities are, first of all, the tangents $x_\alpha^\mu$ and normals $N_i^\mu$ to the worldsheet, $R^\alpha{}_{i\beta j}$ is the spacetime Riemann tensor, suitably contracted with the tangents and normals, $K_{i \alpha\beta}$ is the second fundamental form (extrinsic curvature) of the worldsheet, and $\nabla_\alpha$ the covariant derivative on the worldsheet including the connections on the normal bundle (see appendix~\ref{embed} for details). Furthermore, $\epsilon^{\alpha\beta} = \frac1{\sqrt{-g}} \varepsilon^{\alpha\beta}$ is the Levi-Civita tensor.
We mention that several total derivatives have been dropped in the passage from \eqref{NG:action.NG} to \eqref{NG:action.expand}. 

The field equation of the classical string worldsheet derives from demanding that the first order term \eqref{NG:L1} vanish,
\begin{equation}
\label{NG:eom.bg}
	-K_{i\alpha}{}^\alpha + \frac12 \epsilon^{\alpha\beta}H_{\alpha\beta i} =0~.
\end{equation}
Clearly, the background worldsheet is minimal only in the absence of a background Kalb-Ramond field $H_{\alpha\beta i}$. Equation \eqref{NG:eom.bg} can be simplified, because, in two dimensions,  
\begin{equation}
\label{NG:H.abi}
	H_{\alpha\beta i} = - \frac12 \epsilon_{\alpha\beta} \epsilon^{\gamma\delta} H_{\gamma\delta i}~,
\end{equation}
so that \eqref{NG:eom.bg} becomes 
\begin{equation}
\label{NG:H.abi.2}
	H_{\alpha\beta i} = - \epsilon_{\alpha\beta} K_{i\gamma}{}^\gamma~.
\end{equation}

\subsection{Quadratic action for the fluctuations}

With $\mathcal{L}_1$ vanishing on a classical background, let us turn to $\mathcal{L}_2$.
Substituting \eqref{NG:H.abi.2} into \eqref{NG:L2}, the second term in the brackets cancels against the last term in the parentheses on the first line. Moreover, the last term in the brackets on the second line in \eqref{NG:L2} can be absorbed into the kinetic term by modifying the connection in the normal bundle (see appendix~\ref{embed:structure}) to \cite{Singh:2023olv}
\begin{equation}
\label{NG:mod.conn.norm}
	\tilde{A}_{ij\alpha} = A_{ij\alpha} + \frac12 \epsilon_\alpha{}^\beta H_{\beta i j}
\end{equation}
and, consequently, defining the modified covariant derivative 
\begin{equation}
\label{NG:mod.cov.der}
	\tilde{\nabla}_\alpha \chi^i = \nabla_\alpha \chi^i -\frac12 \epsilon_\alpha{}^\beta H_\beta{}^i{}_j \chi^j 
	= \partial_\alpha \chi^i + \chi^j \tilde{A}_{\alpha j}{}^i~.
\end{equation}

After performing these simplifications, \eqref{NG:L2} becomes
\begin{equation}
\label{NG:L2.simp}
	\mathcal{L}_2 = \frac12 (\tilde{\nabla}^\alpha \chi^i)(\tilde{\nabla}_\alpha \chi_i) + \frac12 \mathcal{M}_{ij}\chi^i \chi^j 
	+ \frac{i}2 \bar{\theta} \left( \Gamma^\alpha \mathcal{D}_\alpha + \epsilon^{\alpha\beta} \geleven\Gamma_\alpha \mathcal{D}_\beta \right) \theta~,
\end{equation}
with the bosonic ``mass'' matrix
\begin{equation}
\label{NG:scal.mass.mat}
	\mathcal{M}_{ij} = - R^\alpha{}_{i\alpha j} -K_{i \alpha\beta} K_j{}^{\alpha\beta} +\frac12 \epsilon^{\alpha\beta} x_\alpha^\mu x_\beta^\nu N_i^\rho N_j^\lambda \nabla_{(\rho} H_{\lambda)\mu\nu} +\frac14 H_{\alpha ki} H^{\alpha k}{}_j~.
\end{equation}

In order to manipulate the fermion terms in \eqref{NG:L2.simp}, we note that the matrices $\Gamma_\alpha$ satisfy 
\begin{equation}
\label{NG:Clifford}
	\Gamma_{(\alpha} \Gamma_{\beta)} = g_{\alpha\beta}~,
\end{equation}
so that, in two dimensions, one has the identity 
\begin{equation}
\label{NG:Gamma.ident}
	\Gamma_\alpha \epsilon^{\alpha\beta} = \gtwo g^{\alpha\beta} \Gamma_\alpha~,  \qquad \gtwo = \frac12 \epsilon_{\alpha\beta} \Gamma^{\alpha\beta}~.
\end{equation}
Thus,
\begin{equation}
\label{NG:ferm.simp}
	\frac{i}2 \left( \Gamma^\alpha \mathcal{D}_\alpha + \epsilon^{\alpha\beta} \geleven\Gamma_\alpha \mathcal{D}_\beta \right) 
	= i \Pi_+ \Gamma^\alpha \mathcal{D}_\alpha~,
\end{equation}
with  $\Pi_+$ being one of the projectors
\begin{equation}
\label{NG:Pi}
	\Pi_\pm = \frac12 \left(1\pm\gtwo\geleven\right)~.
\end{equation}
Spelling out $\mathcal{D}_\alpha$ using \eqref{embed:spin.der.pullback.cov}, we explicitly get
\begin{align}
\notag
	\Pi_+ \Gamma^\alpha \mathcal{D}_\alpha &= \Pi_+ \Gamma^\alpha \left( \nabla_\alpha +\frac12 K_{i\alpha\beta}\Gamma^i \Gamma^\beta + \frac14 H_{\alpha \beta i} \Gamma^\beta\Gamma^i \geleven +\frac18 H_{\alpha ij} \Gamma^{ij} \geleven 
	+\frac18 S\Gamma_\alpha \right)\\
\label{NG:op.ferm}
	&= \Pi_+ \left(\Gamma^\alpha \nabla_\alpha -\frac12 K_{i\alpha}{}^\alpha \Gamma^i - \frac14 H_{\alpha \beta i} \Gamma^{\alpha\beta} \geleven \Gamma^i - \frac18 \geleven H_{\alpha ij} \Gamma^\alpha \Gamma^{ij} +\frac14 \tilde{S} \right)~,
\end{align}
where $\tilde{S}$ has been defined as 
\begin{equation}
\label{NG:S.tilde}
	\tilde{S} = \frac12 \Gamma^\alpha S \Gamma_\alpha~.
\end{equation}
Then, the background field equation \eqref{NG:H.abi.2} and the identities $\Pi_+\geleven = \Pi_+ \gtwo$ and \eqref{NG:Gamma.ident}  further simplify \eqref{NG:op.ferm} to
\begin{align}
\notag
	\Pi_+ \Gamma^\alpha \mathcal{D}_\alpha &= \Pi_+ \left[\Gamma^\alpha \left( \nabla_\alpha -\frac18 \epsilon_\alpha{}^\beta H_{\beta ij} \Gamma^{ij} \right) -\Pi_-  K_{i\alpha}{}^\alpha \Gamma^i +\frac14 \tilde{S} \right]\\
\label{NG:op.ferm.2}
	&= \Pi_+ \left( \Gamma^\alpha \tilde{\nabla}_\alpha +\frac14 \tilde{S} \right)~.
\end{align}
In the last step, we have defined the modified spinor covariant derivative $\tilde{\nabla}_\alpha$ by absorbing $\epsilon_\alpha{}^\beta H_{\beta ij}$ into the connection on the normal bundle, in analogy to the scalars, see \eqref{NG:mod.conn.norm} and appendix~\ref{embed:spinors}. Hence, the scalars and spinors are gauged on the normal bundle in exactly the same fashion.

A short calculation shows that $\tilde{S}$ does not contain terms with odd numbers of $\Gamma^i$'s,
\begin{equation}
\label{NG:S.tilde.expl}
	\tilde{S} = \e{\phi} \left( -\frac12 F_{ij}\Gamma^{ij} \geleven 
	+\frac12 F_{\alpha\beta}\Gamma^{\alpha\beta} \geleven -\frac14 F_{ij\alpha\beta} \Gamma^{ij} \Gamma^{\alpha\beta}
	+\frac1{4!} F_{ijkl} \Gamma^{ijkl} \right)~,
\end{equation}   
and that the projection with $\Pi_+$ leads to the further simplification
\begin{equation}
\label{NG:S.tilde.proj}
	\Pi_+ \tilde{S} = \Pi_+ \e{\phi} \left[\frac1{4!} F_{ijkl} \Gamma^{ijkl} - \frac12 F_{\alpha\beta} \epsilon^{\alpha\beta} -\frac12 \left( F_{ij} - \frac12 F_{ij\alpha\beta} \epsilon^{\alpha\beta} \right) \Gamma^{ij} \gtwo \right]~.
\end{equation} 
Evidently, the eight matrices $\Gamma^i$ appear in the fermion action only through the $SO(8)$ generators $\Gamma^{ij}$. These matrices, which we recall are $32 \times 32$ matrices, split into the direct sum of two $8\times 8$ irreducible representations of $SO(8)$, one left-handed and one-right handed, times a $2\times 2$ unit matrix. Now, the projector $\Pi_+$ selects the $SO(8)$ representation for which $\gnine = \gtwo\geleven=1$. Therefore, we may simply regard $\Pi_+\theta$ as an octet of 2-d spinors, with all matrices being $16\times 16 = (8\cdot 2) \times (8\cdot 2)$ matrices. The matrices $\Gamma_\alpha$ are the usual 2-d Dirac matrices in the $2\times 2$ factor and trival in the $8\times 8$ factor, whereas the matrices $\Gamma^{ij}$ are non-trivial only in the $8\times 8$ factor. 

Thence, the final expression of the second-order action is 
\begin{equation}
\label{NG:L2.final}
	\mathcal{L}_2 = \frac12 (\tilde{\nabla}^\alpha \chi^i)(\tilde{\nabla}_\alpha \chi_i) + \frac12 \mathcal{M}_{ij}\chi^i \chi^j 
	+ i \bar{\theta} \left( \Gamma^\alpha \tilde{\nabla}_\alpha + \frac14 \tilde{S} \right) \theta~, 
\end{equation}
where the spinor $\theta$ now has 16 components (octet of 2-d spinors) and, with some abuse of notation, $\tilde{S}$ has become
\begin{equation}
\label{NG:S.tilde.new}
	\tilde{S} = \e{\phi} \left[\frac1{4!} F_{ijkl} \Gamma^{ijkl} - \frac12 F_{\alpha\beta} \epsilon^{\alpha\beta} -\frac12 \left( F_{ij} - \frac12 F_{ij\alpha\beta} \epsilon^{\alpha\beta} \right) \Gamma^{ij} \gtwo \right]~.
\end{equation} 
For later use, let us provide the following
\begin{equation}
\label{NG:S.tilde.gamma}
	\frac12 \Gamma^\alpha \tilde{S} \Gamma_\alpha = \e{\phi} \left[\frac1{4!} F_{ijkl} \Gamma^{ijkl} - \frac12 F_{\alpha\beta} \epsilon^{\alpha\beta} +\frac12 \left( F_{ij} - \frac12 F_{ij\alpha\beta} \epsilon^{\alpha\beta} \right) \Gamma^{ij} \gtwo \right]
\end{equation} 
and note that  
\begin{equation}
\label{NG:S.tilde.gamma.prop}
	\frac12 \Gamma^\alpha \tilde{S} \Gamma_\alpha \Gamma^\beta = \Gamma^\beta \tilde{S}~.
\end{equation}

We emphasize that, contrary to what has often been stated in the literature, there is no need to gauge fix the kappa symmetry. The projector $\Pi_+$ has taken care of removing the gauge dependent spinor components from the action.

\subsection{Semiclassical effective action and its UV divergences}
\label{NG:anomaly}
The semiclassical path integral over the eight scalars and 2-d Majorana spinors, which live on the background worldsheet, yields the effective action of the superstring for the particular classical background configuration. We will briefly discuss some of its general aspects, in particular the cancellation of the UV divergences. This discussion builds on earlier results \cite{Drukker:2000ep, Drukker:2011za, Aguilera-Damia:2014bqa, Forini:2015mca, Forini:2015bgo, Faraggi:2016ekd, Cagnazzo:2017sny, Forini:2017whz, Aguilera-Damia:2018bam, Medina-Rincon:2019bcc, Gautason:2021vfc, Giombi:2020mhz, Astesiano:2024sgi} and extends them to string worldsheets in arbitrary type-IIA supergravity backgrounds.

Formally, the semiclassical effective action is a combination of functional determinants that are the result of the functional integrals over the bosons and fermions. One ingredient in the functional integration is the Lagrangian \eqref{NG:L2.final}. The overall factor $1/(2\pi\alpha')$ in the action \eqref{NG:action.expand} is irrelevant and can be absorbed into the normalization of the fields. For notational simplicity, we will also drop the tildes. Furthermore, we will switch to Euclidean signature by performing a Wick rotation, such that the relevant second order operators are elliptic.
The other ingredient in the functional integration, which must not be forgotten, is the integral measure. It is defined in terms of the norms of the fluctuation fields. 

Functional determinants generically possess UV divergences, which are elegantly described in the heat kernel method \cite{Gilkey:1995mj, Vassilevich:2003xt}. In 2-d, these may be quadratic, linear and logarithmic in the UV cut-off $\Lambda$.  
As we will explain below in more detail, the quadratic and linear divergences cancel exactly between the bosons and fermions due to the matching numbers of degrees of freedom, as do parts of the logarithmically divergent terms \cite{Giombi:2020mhz}. Other parts of the logarithmically divergent terms cancel by virtue of the supergravity field equations. At the end, there may be a remainder, but we will show that it can be absorbed by an appropriate definition of the integral measure, making the effective action manifestly finite.\footnote{We will not discuss other ways to cancel the logarithmic UV divergence, because they are not very transparent in the static gauge that we use here. Let us only mention that in specific configurations one may adopt a ``2d supersymmetric'' regularization scheme \cite{Giombi:2020mhz}, in which the logarithmic divergence cancels automatically in a spectral representation of the total effective action.}

\emph{Scalars}. The functional integral over the scalars involves the action and norm  
\begin{align}
\label{NG:scalar.action}
	S_{2B} &= \frac12 \int \rmd^2 \sigma \sqrt{g}\, \chi^i \left( - g^{\alpha\beta} \delta_{ij} \nabla_\alpha \nabla_\beta 
	+\mathcal{M}_{ij} \right) \chi^j~,\\
\label{NG:scalar.norm}
	||\chi||^2 &= \int \rmd^2 \sigma \sqrt{g}\, M(\sigma) \chi^i(\sigma) \chi_i(\sigma)~,
\end{align}
where, following \cite{Drukker:2000ep}, we have included an extra local measure factor $M(\sigma)$.\footnote{$M(\sigma)$ may be thought to be the conformal factor that relates the background intrinsic and induced metrics, or it may be related to some more fundamental definition of the functional measure.} This measure factor may be absorbed into the metric by defining
\begin{equation}
\label{NG:scalar.rescale}
	\hat{g}_{\alpha\beta} = M g_{\alpha\beta}~, 
\end{equation}
without rescaling the scalars (2-d scalars naturally have scaling dimension zero). The result is\footnote{The Weyl transformation leaves the Laplacian on scalars form-invariant. Note that the covariant derivatives still contain the (modified) connections on the normal bundle, but they are irrelevant for this discussion.}
\begin{align}
\label{NG:scalar.action.2}
	S_{2B} &= \frac12 \int \rmd^2 \sigma \sqrt{\hat{g}}\, \chi^i \left( - \hat{g}^{\alpha\beta} \delta_{ij} \hat{\nabla}_\alpha \hat{\nabla}_\beta 
	+\hat{\mathcal{M}}_{ij} \right) \chi^j~,  \qquad \hat{\mathcal{M}}_{ij} = M^{-1} \mathcal{M}_{ij}~,\\
\label{NG:scalar.norm.2}
	||\chi||^2 &= \int \rmd^2 \sigma \sqrt{\hat{g}}\, \chi^i \chi_i~.
\end{align}

\emph{Fermions}. A similar procedure must be applied to the fermions. We start with the action and norm
\begin{align}
\label{NG:fermion.action}
	S_{2F} &= i \int \rmd^2 \sigma \sqrt{g}\, \bar{\theta} \left( \Gamma^\alpha \nabla_\alpha +\frac14 S \right) \theta ~,\\
\label{NG:fermion.norm}
	||\theta||^2 &= \int \rmd^2 \sigma \sqrt{g}\, K(\sigma) \bar{\theta} \theta~,
\end{align}
containing a local measure factor $K(\sigma)$. The following field transformations absorb $K(\sigma)$:
\begin{equation}
\label{NG:fermion.rescale}
	\check{g}_{\alpha\beta} = K^2 g_{\alpha\beta}~, \qquad 
	\check{e}_{\alpha}{}^a = K e_{\alpha}{}^a~, 
	\qquad\check{\theta} = K^{-\frac12} \theta~,
\end{equation}
in accordance with the natural scaling dimension $\frac12$ for spinors in 2-d. In terms of the new fields, the action and norm become\footnote{The rescaling of the fermion ensures that the covariant derivative now contains the spin connection with respect to the new zwei-bein $\check{e}_{\alpha}{}^a$.}
\begin{align}
\label{NG:fermion.action.2}
	S_{2F} &= i \int \rmd^2 \sigma \sqrt{\check{g}}\, \bar{\check{\theta}} \left( \check{\Gamma}^\alpha \check{\nabla}_\alpha +\frac14 \check{S} \right) \check{\theta}~,
	\qquad \check{S}=K^{-1}S~,\\
\label{NG:fermion.norm.2}
	||\check{\theta}||^2 &= \int \rmd^2 \sigma \sqrt{\check{g}}\, \bar{\check{\theta}} \check{\theta}~.
\end{align}

We are ready to write down the effective action,
\begin{equation}
\label{NG.effective.action}
	\Gamma =- \ln Z = \ln \frac{\left(\det \hat{\mathcal{O}}_B \right)^{\frac12}}{\left(\det \check{\mathcal{O}}_F\right)^{\frac14}}~,
\end{equation}
where  
\begin{equation}
\label{NG:OB}
	\hat{\mathcal{O}}_B = -\hat{\nabla}^\alpha \hat{\nabla}_\alpha + \hat{\mathcal{M}}~,
\end{equation}
and $\check{\mathcal{O}}_F$ is the ``square'' of the Dirac operator in \eqref{NG:fermion.action.2} \cite{Gilkey:1995mj, Forini:2015mca},
\begin{align}
\notag
	\check{\mathcal{O}}_F &= \left( -\check{\Gamma}^\alpha \check{\nabla}_\alpha + \frac18 \check{\Gamma}^\alpha\check{S}\check{\Gamma}_\alpha \right)\left( \check{\Gamma}^\beta \check{\nabla}_\beta +\frac14 \check{S} \right) \\
\label{NG:OF}
	&= -\check{\nabla}^\alpha\check{\nabla}_\alpha +\frac14 \check{R}^{(2)} -\frac18 \mathcal{A}_{ij\alpha\beta} \Gamma^{ij}\check{\Gamma}^{\alpha\beta} - \frac14 \check{\Gamma}^\alpha\left( \check{\nabla}_\alpha \check{S}\right) 
	+\frac1{32} \check{\Gamma}^\alpha\check{S}\check{\Gamma}_\alpha\check{S}~.
\end{align} 
Here, $\mathcal{A}_{ij\alpha\beta}$ is the (modified) curvature in the normal bundle (recall that we have dropped the tildes just for notation).
The power of the fermionic determinant in \eqref{NG.effective.action} reflects the fact that the integration is over Majorana fermions, not Dirac fermions.

Let us discuss the divergent part of the effective action, $\Gamma_\infty$. The quadratic and linear divergences involving the Seeley coefficients $a_0$ and $a_1$, respectively \cite{Vassilevich:2003xt}, cancel between the bosons and fermions because of the matching numbers of degrees of freedom,
\begin{equation}
\label{NG.number.match}
	\Tr_{\chi} 1 -\frac12 \Tr_{\theta} 1 = 8 - \frac12 16 = 0~, 
\end{equation}
if we also set
\begin{equation}
\label{NG:MK}
	M=K^2\quad \Leftrightarrow \quad\hat{g}_{\alpha\beta} = \check{g}_{\alpha\beta}~.
\end{equation}

The logarithmic divergence is determined by the Seeley coefficient $a_2$ \cite{Drukker:2000ep, Vassilevich:2003xt}, which yields
\begin{align}
\label{NG:Gamma.log}
	\Gamma_\infty &= -\frac1{4\pi} \ln \Lambda \Bigg[ \int \rmd^2\sigma \sqrt{\hat{g}} \Tr_\chi  \left( \frac16 \hat{R}^{(2)} - \hat{\mathcal{M}} \right) + \int \rmd s \sqrt{\hat{\gamma}}\Tr_\chi \left(\frac13 \hat{K}\right) \\
\notag 
	&\quad -\frac12 \int \rmd^2\sigma \sqrt{\check{g}} \Tr_\theta  \left( \frac16 \check{R}^{(2)} - \frac14 \check{R}^{(2)} - \frac1{32} \check{\Gamma}^\alpha\check{S}\check{\Gamma}_\alpha\check{S} \right) -\frac12 \int \rmd s \sqrt{\check{\gamma}}\Tr_\theta \left(\frac13 \check{K}\right) \Bigg]~. 
\end{align} 
We have included the boundary terms with the appropriate metrics and extrinsic curvatures. In the fermionic terms (second line), we have omitted the manifestly traceless potential terms from \eqref{NG:OF} and kept separate the curvature contribution from the potential (factor $-\frac14$) from the topological contribution (factor $\frac16$). 

After rewriting \eqref{NG:Gamma.log} in terms of the induced metric $g_{\alpha\beta}$ undoing the transformations \eqref{NG:scalar.rescale} and \eqref{NG:fermion.rescale}, it is easy to see that the dependence on the scale factors $M$ and $K$ cancels between the topological curvature terms and the boundary terms \cite{Drukker:2000ep}, both for the bosonic and fermionic determinants.\footnote{This even holds without imposing \eqref{NG:MK}.} In turn, the topological curvature terms and the boundary terms cancel between the bosons and fermions, again because of \eqref{NG.number.match}. These cancellations appear to be universal \cite{Giombi:2020mhz}. We are left with the contributions from the potential terms
\begin{equation}
\label{NG:Gamma.log.2}
	\Gamma_\infty = -\frac1{4\pi} \ln \Lambda \int \rmd^2\sigma \sqrt{g} \left[ - \mathcal{M}^i{}_i + 
	2 R^{(2)} -4 \nabla^\alpha \nabla_\alpha \ln K +\frac1{64}  \Tr_\theta \left( \Gamma^\alpha S\Gamma_\alpha S \right) \right]~.
\end{equation}
Note that the scale factor $K$ arises from the curvature contribution in the fermionic potential, the same that leads to the $2R^{(2)}$.

To further simplify \eqref{NG:Gamma.log.2}, let us start with the contribution from the fermionic mass term. From \eqref{NG:S.tilde.new} and \eqref{NG:S.tilde.gamma} one has
\begin{align}
\notag
	\frac12 \Gamma^\alpha S\Gamma_\alpha S	&= \e{2\phi} \left[\frac1{4!} F_{ijkl} \Gamma^{ijkl} - \frac12 F_{\alpha\beta} \epsilon^{\alpha\beta} +\frac12 \left( F_{ij} - \frac12 F_{ij\alpha\beta} \epsilon^{\alpha\beta} \right) \Gamma^{ij} \gtwo \right]\\
\notag &\quad \times 
	\left[\frac1{4!} F_{ijkl} \Gamma^{ijkl} - \frac12 F_{\alpha\beta} \epsilon^{\alpha\beta} -\frac12 \left( F_{ij} - \frac12 F_{ij\alpha\beta} \epsilon^{\alpha\beta} \right) \Gamma^{ij} \gtwo \right] \\
\label{NG:SS} &=
	\e{2\phi} \Bigg[ \frac1{4!} F_{ijkl}F^{ijkl} - \frac14 F_{ij\alpha\beta}F^{ij\alpha\beta} +\frac12 F_{ij}F^{ij} 
	-\frac12 F_{\alpha\beta}F^{\alpha\beta} \\
\notag &\quad -\frac12 \epsilon_{\alpha\beta} \left( \frac1{(4!)^2} \epsilon^{\alpha\beta i_1\cdots i_8} F_{i_1\cdots i_4} 
	F_{i_5\cdots i_8} + F^{ij\alpha\beta} F_{ij} \right) +\cdots \Bigg]~,
\end{align}
where the ellipses stand for traceless terms. Hence,
\begin{align}
\notag 
	\Tr_\theta \left(\frac1{32} \Gamma^\alpha S\Gamma_\alpha S\right) &=
	 \e{2\phi} \Bigg[ \frac1{4!} F_{ijkl}F^{ijkl} - \frac14 F_{ij\alpha\beta}F^{ij\alpha\beta} +\frac12 F_{ij}F^{ij} 
	-\frac12 F_{\alpha\beta}F^{\alpha\beta} \\
\notag &\quad 
	-\frac12 \epsilon_{\alpha\beta} \left( \frac1{(4!)^2} \epsilon^{\alpha\beta i_1\cdots i_8} F_{i_1\cdots i_4} 
	F_{i_5\cdots i_8} + F^{ij\alpha\beta} F_{ij} \right) \Bigg] \\
\label{NG:SS.trace}
	&= - \e{2\phi} x^{\alpha \mu} x_\alpha^\nu \left[ \left| F_4\right|^2_{\mu\nu} +\left| F_2\right|^2_{\mu\nu} -\frac12 g_{\mu\nu} \left(\left| F_4\right|^2 + \left| F_2\right|^2 \right) \right] \\
\notag &\quad 
	-\frac12 \epsilon_{\alpha\beta} x^\alpha_\mu x^\beta_\nu  \left( \frac1{(4!)^2} \epsilon^{\mu\nu \rho_1\cdots \rho_8} F_{\rho_1\cdots \rho_4} F_{\rho_5\cdots \rho_8} + F^{\rho\lambda\mu\nu} F_{\rho\lambda} \right) \Bigg]~,
\end{align}
where we have used the compact notation \eqref{sugra:F2.abbrev}. Then, using the supergravity field equations, in particular \eqref{sugra:einstein} and \eqref{sugra:Maxwell.h3.coord}, \eqref{NG:SS.trace} becomes  
\begin{align}
\label{NG:Tr.SS}
	\Tr_\theta \left( \frac1{32}\Gamma^\alpha S\Gamma_\alpha S \right) 
	&= - x^{\alpha \mu}x_\alpha^\nu \left( 2 R_{\mu\nu} +4 \nabla_\mu \nabla_\nu \phi -\frac12 H_{\mu\rho\lambda} H_\nu{}^{\rho\lambda} \right)	\\
\notag &\quad 
	+ \epsilon^{\alpha\beta} x_\alpha^\mu x_\beta^\nu \left[ \nabla^\rho H_{\rho\mu\nu} -2(\nabla_\rho \phi) H^\rho{}_{\mu\nu} \right]~.
\end{align}

Turning back to \eqref{NG:Gamma.log.2}, consider $\mathcal{M}^i{}_i$ from \eqref{NG:scal.mass.mat}, 
\begin{align}
\notag
	\mathcal{M}^i{}_i &= - R^{\alpha i}{}_{\alpha i} -K^i{}_{\alpha\beta} K_i{}^{\alpha\beta} +\frac12 \epsilon^{\alpha\beta} x_\alpha^\mu x_\beta^\nu N^{i\rho} N_i^\lambda \nabla_{(\rho} H_{\lambda)\mu\nu} +\frac14 H_{\alpha ij} H^{\alpha ij}\\
\label{NG:Mii}
	&= -R^\alpha{}_\alpha +R^{(2)} - K^{i\alpha}{}_\alpha K_{i\beta}{}^\beta+
	\frac12 \epsilon^{\alpha\beta} x_\alpha^\mu x_\beta^\nu \nabla_{\rho} H^\rho{}_{\mu\nu}
	+\frac14 H_{\alpha ij} H^{\alpha ij}~,
\end{align}
where we have used some geometric identities from appendix~\ref{embed}. 
After combining \eqref{NG:Tr.SS} with \eqref{NG:Mii} and using again some geometric identities, one finds for the combination that appears in \eqref{NG:Gamma.log.2}
\begin{align}
\notag
	\mathcal{M}^i{}_i &-\frac1{64} \Tr_\theta \left( \Gamma^\alpha S\Gamma_\alpha S \right)\\
\notag
	&=  R^{(2)} - K^{i\alpha}{}_\alpha K_{i\beta}{}^\beta -\frac12 H_{\alpha\beta i} H^{\alpha\beta i} 
	+2 \nabla^\alpha \nabla_\alpha \phi - 2 K_{i\alpha}{}^\alpha \nabla^i \phi 
	+ \epsilon^{\alpha\beta} (\nabla^\mu \phi) H_{\mu\alpha\beta} \\
\label{NG:M.diff}
	&= R^{(2)} +2 \nabla^\alpha \nabla_\alpha \phi~,
\end{align}
while all the other terms have cancelled by virtue of the background equation \eqref{NG:H.abi.2}.

Therefore, substituting \eqref{NG:M.diff} into \eqref{NG:Gamma.log.2}, our final result for $\Gamma_\infty$ is
\begin{equation}
\label{NG:Gamma.log.3}
	\Gamma_\infty = -\frac1{4\pi} \ln \Lambda \int \rmd^2\sigma \sqrt{g} \left( 
	R^{(2)}  -2 \nabla^\alpha \nabla_\alpha \phi - 4 \nabla^\alpha \nabla_\alpha \ln K \right)~.
\end{equation}
It is obvious that this can be set to zero by a suitable choice of $K$. For example, for a metric in the conformal form
$g_{\alpha\beta}=\e{2\rho}\eta_{\alpha\beta}$, one would set $K = \e{-\frac12 (\rho+\phi)}$. Together with \eqref{NG:MK}, this implies that the ``preferred'' norms \eqref{NG:scalar.norm} and \eqref{NG:fermion.norm}, for the calculation in static gauge, are
\begin{equation}
\label{NG:preferred.norms}
	||\chi||^2 = \int \rmd^2 \sigma\, g^{\frac14} \e{-\phi} \chi^i \chi_i~,\qquad 
	||\theta||^2 = \int \rmd^2 \sigma\, g^{\frac38} \e{-\frac12 \phi} \bar{\theta}\theta~.
\end{equation}

We refrain from considering the finite part of $\Gamma$, which must be calculated case by case. It would be certainly interesting to work out the implications of \eqref{NG:preferred.norms} for the finite part of $\Gamma$ for specific cases in the literature. 
 
\section{Superstring fluctuations in the Polyakov formulation}
\label{Polyakov}

The standard semiclassical treatment of superstring fluctuations in the Polyakov formulation has been described in a manifestly covariant form, but without the Kalb-Ramond field, in \cite{Forini:2015mca}. It is entirely equivalent to the semiclassical treatment in the Nambu-Goto formulation. The addition of the Kalb-Ramond field does not add anything new in principle. In the first part of this section, we will limit ourselves to establish this equivalence for the general case. 
In the second part, we motivate and describe a non-standard semiclassical treatment.

\subsection{Standard semiclassical treatment}

In the Polyakov formalism, the auxiliary metric appears as an additional field. Its classical background value, according to the field equation \eqref{NG:h.eom}, is conformal to the background induced metric. 
We introduce the fluctuations of the auxiliary metric by writing $h^{\alpha\beta}\to h^{\alpha\beta}+ \delta h^{\alpha\beta}$.
Then, up to second order in $\delta h^{\alpha\beta}$, the combination $\sqrt{-h}h^{\alpha\beta}$ can be expressed as
\begin{equation}
\label{P:unih.expand}
	\sqrt{-h} h^{\alpha\beta} \to \unih^{\alpha\beta} + \delta \unih^{\alpha\beta} 
	+ \frac14 \unih^{\alpha\beta} \left(\delta \unih^{\gamma\delta} \unih_{\delta\varepsilon} \delta\unih^{\varepsilon\phi}  \unih_{\phi\gamma} \right) + \mathcal{O}(\delta \unih^3)~,
\end{equation}
where
\begin{equation}
\label{P:unih.def}
	\unih^{\alpha\beta} = \sqrt{-h} h^{\alpha\beta}
\end{equation}
is the unimodular part ($\det \unih^{\alpha\beta}=-1$) of the background auxiliary metric, $\unih_{\alpha\beta}$ its inverse, and $\delta \unih^{\alpha\beta}$ denotes the traceless combination
\begin{equation}
\label{P:delta.unih.def}
	\delta \unih^{\alpha\beta} = \sqrt{-h} \left( 1 -\frac12 \sqrt{-h}\,\delta h^{\epsilon\phi} h_{\epsilon\phi} \right) 
	\left( \delta h^{\alpha\beta}-\frac12 h^{\alpha\beta} \delta h^{\gamma\delta} h_{\gamma\delta} \right)~.
\end{equation}
The fact that the background induced and auxiliary metrics are conformal to each other is now simply expressed as 
\begin{equation}
\label{P:unigh.background}
	\unih^{\alpha\beta} = \unig^{\alpha\beta}\equiv \sqrt{-g} g^{\alpha\beta}~.
\end{equation}

Let us consider the expansion of the action \eqref{intro:GS.action} up to second order in the fluctuations. For simplicity, we will work in the static gauge setting the longitudinal fluctuations to zero. The remark made in the previous section applies also here. If one keeps the longitudinal fluctuations, one would find that their contributions to the action would eventually drop out or reduce to boundary terms upon using the background field equations.
It is easy to see that the background and first-order terms of the Lagrangian remain exactly as in \eqref{NG:L0} and \eqref{NG:L1}, respectively. In particular, the background field equation is again \eqref{NG:H.abi.2}. 
In the second-order Lagrangian, the fermionic terms also remain as in \eqref{NG:L2}, whereas the bosonic terms are now 
\begin{align}
\label{P:L2bos}
	\mathcal{L}_{2B} &= \frac12 g^{\alpha\beta} (\nabla_\alpha \chi^i)(\nabla_\beta \chi_i) - \frac12 g^{\alpha\beta}
	\left( R_{\alpha i\beta j} - K_{i\alpha\gamma} K_{j\beta}{}^\gamma \right)  \chi^i\chi^j \\
\notag 
	&\quad + \frac12 \epsilon^{\alpha\beta} \left[ \frac12 x_\alpha^\mu x_\beta^\nu N_i^\rho N_j^\lambda \left(\nabla_\rho H_{\lambda\mu\nu} \right) \chi^i \chi^j + K_{i\alpha}{}^\gamma H_{\beta \gamma j} \chi^i \chi^j  +H_{\alpha i j} \chi^j \nabla_\beta \chi^i \right] \\
\notag &\quad	 
	- \frac1{\sqrt{-g}} \delta \unih^{\alpha\beta} K_{i\alpha\beta} \chi^i
	+\frac14 \delta \unih^{\alpha\beta} \unig_{\beta\gamma} \delta\unih^{\gamma\delta} \unig_{\delta\alpha}~.
\end{align}
The easiest way to proceed is to decouple the auxiliary field $\delta\unih^{\alpha\beta}$ from the other fields by performing the shift \cite{Forini:2015mca}
\begin{equation}
\label{P:delta.h.shift}
	\delta \unih^{\alpha\beta} \to \delta\unih^{\alpha\beta} +\left( 2 \unig^{\gamma(\alpha} K_{i\gamma}{}^{\beta)} - \unig^{\alpha\beta} K_{i\gamma}{}^\gamma \right) \chi^i~.
\end{equation}
This produces a term quadratic in $\delta\unih^{\alpha\beta}$, which allows to trivially integrate out $\delta\unih^{\alpha\beta}$,  and an additional term which, in combination with the rest of \eqref{P:L2bos}, reproduces precisely the bosonic terms of \eqref{NG:L2}. This shows the equivalence of the semiclassical treatment in the Polyakov and Nambu-Goto formulations. This is no surprise, because the shift \eqref{P:delta.h.shift} simply incorporates the solution of the field equation \eqref{NG:h.eom} at first order.

\subsection{Non-standard semiclassical treatment}
\label{P:nonstandard}

As we mentioned in the introduction, there are applications, in which the standard semiclassical treatment fails, because, loosely speaking, some fluctuations become large. In order to avoid such a situation, one may attempt to set up the semiclassical calculation by expanding only the bosonic and fermionic coordinates around the classical background solution, but keeping the auxiliary metric $h^{\alpha\beta}$ arbitrary, \ie classically off-shell. (We only consider backgrounds with vanishing fermions, so that the fermionic terms in the action are already of second order.) This would allow the difference between the induced and auxiliary metrics to be ``large''. More precisely, one should consider the unimodular parts of these metrics,
\begin{equation}
\label{P:hg.unimodular}
	\unig^{\alpha\beta} = \sqrt{-g} g^{\alpha\beta}~,\qquad \unih^{\alpha\beta} = \sqrt{-h} h^{\alpha\beta}~,
\end{equation}
which are now independent of each other. The idea, then, would be to first perform the Gaussian integrals over the coordinate fluctuations, which should result in an effective action $\Gamma[\unih^{\alpha\beta}]$, and then find the configuration $\unih^{\alpha\beta}$ that minimizes the effective action. We anticipate that we are not able to provide a first-principle treatment of this progamme. In the rest of this section, we will describe in detail the obstacles one encounters along the way.
Under certain circumstances, which are void of some of these obstacles, but nevertheless correspond to physically interesting situations, we are able to construct a self-consistent solution. This will be the subject of section~\ref{Hage}.

Before starting, let us mention a few useful identities. Because both $\unig_{\alpha\beta}$ and $\unih_{\alpha\beta}$ are unimodular, they satisfy 
\begin{equation}
\label{P:hg.ident}
	\varepsilon_{\alpha\gamma}\varepsilon_{\beta\delta} 
	= -\unih_{\alpha\beta} \unih_{\gamma\delta} + \unih_{\alpha\delta} \unih_{\beta\gamma}  
	= -\unig_{\alpha\beta} \unig_{\gamma\delta} + \unig_{\alpha\delta} \unig_{\beta\gamma}~,
\end{equation}
from which one may derive 
\begin{equation}
\label{P:h.down}
	\unih_{\alpha\beta} = 2\Omega \unig_{\alpha\beta} -\unig_{\alpha\gamma} \unih^{\gamma\delta}\unig_{\delta\beta}
\end{equation}  
as well as\footnote{This is just the property of $2\times2$ matrices $\Tr(A^2) = (\Tr A)^2 - 2 \det A$.}
\begin{equation}
\label{P:ghgh}
	\unig^{\alpha\beta} \unih_{\beta\gamma}\unig^{\gamma\delta} \unih_{\delta\alpha} = 4\Omega^2 -2~, 
\end{equation}
where $\Omega$ stands for\footnote{We remark that the equality $\unig^{\alpha\beta} \unih_{\alpha\beta}= \unih^{\alpha\beta} \unig_{\alpha\beta}$ has nothing to do with raising or lowering indices, but follows from the fact that the matrix $\unig^{\alpha\beta} \unih_{\beta\gamma}$ is the inverse of $\unih^{\alpha\beta} \unig_{\beta\gamma}$. A $2\times 2$ matrix with unit determinant has the same trace as its inverse.}
\begin{equation}
\label{P:Omega.def}
	\Omega = \frac12 \unih^{\alpha\beta} \unig_{\alpha\beta} = \frac12 \unig^{\alpha\beta}\unih_{\alpha\beta}~.
\end{equation}

Let us now expand the action \eqref{intro:GS.action} up to quadratic order in the coordinate fluctuations. As before, we will work in the static gauge setting the longitudinal fluctuations to zero. 
One easily finds\footnote{Note that the integral measure is now included in the Lagrangians.}
\begin{equation}
\label{P:action.expand}
	S_P = -\frac1{2\pi\alpha'} \int  \rmd^2\sigma \left(\mathcal{L}_0 + \mathcal{L}_1 + \mathcal{L}_2 +\cdots \right)  
\end{equation}
with 
\begin{align}
\label{P:L0}
	\mathcal{L}_0 &= \frac12 \unih^{\alpha\beta} g_{\alpha\beta} + \frac12 \varepsilon^{\alpha\beta} b_{\alpha\beta}~,\\
\label{P:L1}
	\mathcal{L}_1 &=  \left( - \unih^{\alpha\beta} K_{i\alpha\beta} + \frac12\varepsilon^{\alpha\beta}H_{\alpha\beta i} \right) \chi^i~,\\
\label{P:L2}
	\mathcal{L}_2 &= \frac12 \unih^{\alpha\beta} (\nabla_\alpha \chi^i)(\nabla_\beta \chi_i) - \frac12  \unih^{\alpha\beta}
	\left( R_{\alpha i\beta j} - K_{i\alpha\gamma} K_{j\beta}{}^\gamma \right)  \chi^i\chi^j \\
\notag 
	&\quad + \frac12 \varepsilon^{\alpha\beta} \left[ \frac12 x_\alpha^\mu x_\beta^\nu N_i^\rho N_j^\lambda \left(\nabla_\rho H_{\lambda\mu\nu} \right) \chi^i \chi^j + K_{i\alpha}{}^\gamma H_{\beta \gamma j} \chi^i \chi^j  +H_{\alpha i j} \chi^j \nabla_\beta \chi^i \right] \\
\notag &\quad	 
	+ \frac{i}2 \bar{\theta} \left( \unih^{\alpha\beta} +\varepsilon^{\alpha\beta} \geleven \right) \Gamma_\alpha \mathcal{D}_\beta \theta~.
\end{align}
Let us stress that we have not introduced fluctuations of $\unih^{\alpha\beta}$, which remains as the full field.
Moreover, we recall our convention that bulk tensors with worldsheet or normal indices simply mean the appropriate pull-backs (using $x_\alpha^\mu$ and $N^\mu_i$) onto the worldsheet or the normal bundle, respectively. Moreover, raising or lowering of worldsheet indices is always done with the induced metric, $g_{\alpha\beta}$, with the exception of the auxiliary metric, for which $\unih_{\alpha\beta}$ is the inverse of $\unih^{\alpha\beta}$.

First, we have to deal with the first-order Lagrangian \eqref{P:L1}. Now, one cannot argue that its vanishing coincides with a background field equation, because it contains the full field $\unih^{\alpha\beta}$. Rather, the background configuration should continue to satisfy the same equation as in the Nambu-Goto formulation, namely \eqref{NG:eom.bg}. Using it, \eqref{P:L1} becomes 
\begin{equation}
\label{P:L1.2}
	\mathcal{L}_1 = \left(\unig^{\alpha\beta}-\unih^{\alpha\beta}\right) K_{i\alpha\beta}\chi^i~,
\end{equation}
which does not vanish in general. Absorbing it by a shift of $\chi^i$, as one would do in standard Gaussian integration, is not an option. In fact, writing the bosonic terms formally as 
$$ b_i\chi^i + \frac12 \chi^i O_{ij} \chi^j = \frac12 (\chi^i +c^i) O_{ij}(\chi^j+c^j) - \frac12 c^i O_{ij} c^j~,\qquad c^i = b_j(O^{-1})^{ji}~,$$
this would not only involve the inverse of the differential operator $O_{ij}$, but also a finite shift $c^i$, whereas $\chi^i$ is a fluctuation. One should not be allowed to shift fluctuations by a finite amount.

An alternative is to perform a variable redefinition of $\unih^{\alpha\beta}$ that is formally identical to the expansion \eqref{P:unih.expand} with 
\begin{equation}
\label{P:delta.unih.sol}
	\delta\unih^{\alpha\beta} =\left( 2 \unih^{\gamma(\alpha} K_{i\gamma}{}^{\beta)}  
	- \frac{\unig^{\alpha\beta} + \unih^{\alpha\beta}}{1+\Omega} K_{i\gamma}{}^\gamma\right) \chi^i~.
\end{equation}
Applying this shift to $\mathcal{L}_0$, the first-order term in $\mathcal{L}_1$ is cancelled. Moreover, the second-order terms of the redefinition \eqref{P:unih.expand}, together with the first-order shift applied to $\mathcal{L}_1$, produce a number of second-order terms, which contribute to the scalar mass terms in $\mathcal{L}_2$, in a similar way as was discussed for the shift \eqref{P:delta.h.shift}. In fact, in the standard setup with $\unih^{\alpha\beta}=\unig^{\alpha\beta}$, \eqref{P:delta.unih.sol} coincides with \eqref{P:delta.h.shift}. 
These second order terms are 
\begin{align}
\notag 	&-\delta \unih^{\alpha\beta}K_{i\alpha\beta}\chi^i + \frac18 \unih^{\alpha\beta} g_{\alpha\beta}
	\left(\delta \unih^{\gamma\delta}\unih_{\delta\epsilon}\delta \unih^{\epsilon\phi} \unih_{\phi_\gamma} \right) \\
\label{P:L2B.2}
	& \qquad = \left( - \unih^{\alpha\beta} K_{i\alpha\gamma} K_{j\beta}{}^\gamma  + \frac12 \unig^{\alpha\beta} K_{i\alpha\beta} K_{j\gamma}{}^\gamma  \right) \chi^i \chi^j \\
\notag
	&\qquad \quad 
	+\frac12 \unig_{\alpha\beta} 
	\left( K_{i\gamma}{}^\alpha \unih^{\beta\gamma} -\frac12 \unih^{\alpha\beta} K_{i\gamma}{}^\gamma \right) 
	\left( \unig^{\delta\epsilon} -\Omega \unih^{\delta\epsilon} \right) K_{j\delta\epsilon} \chi^i \chi^j \\
\notag
	&\qquad \quad 	
	+\frac14 \unig_{\alpha\beta} \left( \unig^{\alpha\beta} - \Omega \unih^{\alpha\beta} \right) 
	\left[ \unih^{\gamma\delta} K_{i\gamma\delta} K_{j\epsilon}{}^\epsilon - 2 \unih^{\gamma\delta} K_{i\gamma\epsilon} K_{j\delta}{}^\epsilon +\frac1{(1+\Omega)^2}\unig^{\gamma\delta} K_{i\gamma\delta} K_{j\epsilon}{}^\epsilon 
	\right] \chi^i \chi^j 
\end{align}

Furthermore, the bosonic terms in \eqref{P:L2} can be manipulated in a similar way as was done in the case of the Nambu-Goto formulation. First, the term containing $(\nabla_\alpha \chi^i) \chi^j$ is absorbed into the connection on the normal bundle by defining
\begin{equation}
\label{P:mod.conn.norm}
	\tilde{A}_{ij\alpha} = A_{ij\alpha} - \frac12 \unih_{\alpha\beta}\varepsilon^{\beta\gamma} H_{\gamma i j} = 
	A_{ij\alpha} - \frac12 \varepsilon_{\alpha\beta}\unih^{\beta\gamma} H_{\gamma i j}
\end{equation}
and 
\begin{equation}
\label{P:mod.cov.der}
	\tilde{\nabla}_\alpha \chi^i = \partial_\alpha \chi^i + \tilde{A}^i{}_{j\alpha} \chi^j~.
\end{equation}
Second, the term containing $H_{\beta \gamma j}$ (on the second line) is rewritten using the background field equation \eqref{NG:eom.bg}. These two steps yield
\begin{align}
\notag 
	&\frac12 \unih^{\alpha\beta} (\nabla_\alpha \chi^i)(\nabla_\beta \chi_i) - \frac12  \unih^{\alpha\beta}
	\left( R_{\alpha i\beta j} - K_{i\alpha\gamma} K_{j\beta}{}^\gamma \right)  \chi^i\chi^j \\
\notag 
	& + \frac12 \varepsilon^{\alpha\beta} \left[ \frac12 x_\alpha^\mu x_\beta^\nu N_i^\rho N_j^\lambda \left(\nabla_\rho H_{\lambda\mu\nu} \right) \chi^i \chi^j + K_{i\alpha}{}^\gamma H_{\beta \gamma j} \chi^i \chi^j  +H_{\alpha i j} \chi^j \nabla_\beta \chi^i \right] \\
\label{P:L2B.1}
	&\qquad \qquad = \frac12 \unih^{\alpha\beta} (\tilde{\nabla}_\alpha \chi^i)(\tilde{\nabla}_\beta \chi_i) 
	+\frac12 \unih^{\alpha\beta} \left( -R_{\alpha i \beta j} + K_{i\alpha\gamma} K_{j\beta}{}^\gamma 
	+ \frac14  H_{\alpha ik} H_{\beta j}{}^k \right) \chi^i \chi^j\\
\notag
	& \qquad \qquad \quad 
	-\frac12\unig^{\alpha\beta} K_{i\alpha\beta} K_{j\gamma}{}^\gamma \chi^i \chi^j
	+ \frac14 \varepsilon^{\alpha\beta}x_\alpha^\mu x_\beta^\nu N_i^\rho N_j^\lambda \left(\nabla_\rho H_{\lambda\mu\nu} \right) \chi^i \chi^j~.
\end{align}
The bosonic second-order Lagrangian $\mathcal{L}_{2B}$ is now the sum of \eqref{P:L2B.1} and \eqref{P:L2B.2}. 

The same mechanism of field redefinitions, however, renders the quadratic action ambiguous. 
In fact, one may perform other shifts of $\unih^{\alpha\beta}$ of the form \eqref{P:delta.h.shift} with  
\begin{align}
\label{P:h.shift.generic1}
	\delta \unih^{\alpha\beta} &= \left( 2 \unih^{\gamma(\alpha} A_\gamma{}^{\beta)}  
		- \frac{\unig^{\alpha\beta} + \unih^{\alpha\beta}}{1+\Omega} A_\gamma{}^\gamma\right) \\
\intertext{or}
\label{P:h.shift.generic2}
	\delta \unih^{\alpha\beta} &= \left( \unig^{\alpha\beta} - \Omega \unih^{\alpha\beta} \right) B~,
\end{align}
where $A_\alpha{}^\beta$ and $B$ are arbitrary second-order expressions, bosonic or fermionic. It is quite easy to convince oneself that by using such shifts the entire second-order Lagrangian can be transformed into the second-order Lagrangian of the Nambu-Goto formulation \eqref{NG:L2.final}, or into other forms. Clearly, this is not what we had in mind. Moreover, such shifts may induce non-trivial Jacobian determinants, which are difficult to keep trace of, because we have not defined the integration measures to begin with. Therefore, we will not further pursue this path.

Two additional problems arise in the fermionic sector. One minor issue is that the generalized covariant derivative $\mathcal{D}_\alpha$ naturally contains the spin connections relative to the induced metric, not the auxiliary metric. This can simply be addressed by rewriting it as the sum of the spin connections for the auxiliary metric and the difference between the two connections, and the difference considered to contribute some mass term. The other problem has to do with $\kappa$ symmetry and is more severe.
The fermionic part of \eqref{P:L2} can be written in a concise form by first introducing the projectors 
\begin{equation}
\label{P:projectors}
	P_\pm^{\alpha\beta} = \frac12 \left( \unih^{\alpha\beta} \pm \varepsilon^{\alpha\beta}\geleven \right)~,
\end{equation}
which satisfy
\begin{equation}
\label{P:proj.prop}
	P_\pm^{\alpha\beta} \unih_{\beta\gamma} P_\pm^{\gamma\delta} = P_\pm^{\alpha\delta}~,\qquad
	P_\pm^{\alpha\beta} \unih_{\beta\gamma} P_\mp^{\gamma\delta} = 0~.
\end{equation}
These projectors are, of course, those involved in the $\kappa$ symmetry transformations \cite{Sorokin:1999jx}
\begin{equation}
\label{P:k.sym}
	\delta_\kappa \theta = i\Gamma_m E_\alpha^m \kappa^\alpha~,
\end{equation} 
where $E_\alpha^m$ is the pull-back of the super vielbein that is compatible with the full metric in superspace
\begin{equation}
\label{P:vielbein.E}
	E_\alpha^m = x_\alpha^\mu E_\mu^m~,\qquad E_\mu^m E_\nu^n \eta_{mn} = G_{\mu\nu}~.
\end{equation}
Furthermore, the parameters $\kappa^\alpha$ in \eqref{P:k.sym} must satisfy the condition
\begin{equation}
\label{P:k.cond}
	P_+^{\alpha\beta} h_{\beta\gamma} \kappa^\gamma =0~. 
\end{equation}

Now, the fermionic term in \eqref{P:L2} is simply
\begin{equation}
\label{P:L2f.1}
	\mathcal{L}_{2F} = i \bar{\theta} P_+^{\alpha\beta} \Gamma_\alpha \mathcal{D}_\beta \theta
	= i \bar{\theta} \Gamma_\alpha P_-^{\alpha\beta} \mathcal{D}_\beta \theta~. 
\end{equation}
An important fact to remember is that the matrices $\Gamma_\alpha= x_\alpha^\mu \Gamma_\mu$ satisfy
\begin{equation}
\label{P:Gamma.anticomm}
	\Gamma_{(\alpha}\Gamma_{\beta)} = g_{\alpha\beta}~,
\end{equation}
which implies \eqref{NG:Gamma.ident}, or 
\begin{equation}
\label{P:Gamma.ident}
	\Gamma_\alpha \varepsilon^{\alpha\beta} = \gtwo \Gamma_\alpha \unig^{\alpha\beta}~. 
\end{equation}
Therefore, the combination $\Gamma_\alpha P_-^{\alpha\beta}$ in \eqref{P:L2f.1} can be rewritten as
\begin{equation} 
\label{P:L2f.comb}
	\Gamma_\alpha P_-^{\alpha\beta} = \frac12 \Gamma_\alpha \left(\unih^{\alpha\beta} + \unig^{\alpha\beta} \gtwo\geleven \right)
	= \frac12 \Gamma_\alpha \left[ \left(\unih^{\alpha\beta} + \unig^{\alpha\beta} \right) \Pi_+ 
	+ \left(\unih^{\alpha\beta} - \unig^{\alpha\beta} \right) \Pi_- \right]~,
\end{equation}
with the projectors $\Pi_\pm$ that were defined in \eqref{NG:Pi}.\footnote{These projectors are independent of the worldsheet metric.}
So, with $\unih^{\alpha\beta}$ off-shell, there is no projector that removes half of the fermionic components. 
In addition, recall that \eqref{P:k.cond} implies that $\kappa^\alpha$ has 32 independent components. Only if $h^{\alpha\beta}$ is on-shell, which here means it satisfies
\begin{equation}
\label{P:h.onshell}
	\left( E_\alpha^m E_\beta^n -\frac12 h_{\alpha\beta} h^{\gamma\delta} E_\gamma^m E_\delta^n\right) \eta_{mn} =0~,
\end{equation}
then the symmetry transformation \eqref{P:k.sym} acts on only 16 components of $\theta$.

In order to address this problem, we must consider, instead of \eqref{P:L2f.1}, the fermionic action
\begin{equation}
\label{P:L2f.2}
	\mathcal{L}_{2F} = i \bar{\theta} P_+^{\alpha\beta} E_\alpha^m \Gamma_m \mathcal{D}_\beta \theta~. 
\end{equation}
Similarly, those terms in $\mathcal{D}_\beta$ that involve $\Gamma_\beta= x_\beta^\mu \Gamma_\mu$ (only those with the same index $\beta$) should be replaced by $E_\beta^m \Gamma_m$. 
Note that, up to second order in the fermions,
\begin{equation}
\label{P:E.2.ord}
	E_\alpha^m = x_\alpha^\mu e_\mu^m + \frac12 i \bar\theta \Gamma^m \mathcal{D}_\alpha \theta~,
\end{equation}
where $e_\mu^m$ is the purely bosonic vielbein. Therefore, \eqref{P:L2f.2} agrees to second order with \eqref{P:L2f.1} and contains some, but not all, of the fourth-order fermionic terms listed in \cite{Wulff:2013kga}.
The problem reappears, if we insert the expansion \eqref{P:E.2.ord} into \eqref{P:L2f.2} and simply drop the fourth-order terms, which would result in \eqref{P:L2f.1}. Rather, to get a purely second-order Lagrangian for a 16-component spinor, one should replace $E_\alpha^m$ in \eqref{P:L2f.2} by a formally zeroth-order quantity that satisfies \eqref{P:h.onshell}. Such a quantity is easy to find and is given, up to a conformal factor that can always be absorbed by a conformal rescaling of the fermions, by
\begin{equation}
\label{P:E.classical}
	E_\alpha^m \to \bar{E}_\alpha^m = \unie_\alpha{}^a g_a{}^\beta x_\beta^\mu E_\mu^m~,   
\end{equation} 
where $\unie_\alpha{}^a\unie_\beta{}^b \eta_{ab} = \unih_{\alpha\beta}$, but $g_a{}^\alpha g_b{}^\beta g_{\alpha\beta}= \eta_{ab}$.
Then, the matrices 
\begin{equation}
\label{P:gamma.new.def}
	\gamma_\alpha = \bar{E}_\alpha^m \Gamma_m
\end{equation}
satisfy the relations 
\begin{equation}
\label{P:gamma.new.Clifford}
	\gamma_{(\alpha} \gamma_{\beta)} = \unih_{\alpha\beta}~, \qquad 
	\gamma_\alpha \varepsilon^{\alpha\beta} = \gtwo \gamma_\alpha \unih^{\alpha\beta}~,
\end{equation}
so that the action \eqref{P:L2f.2}, after the replacement \eqref{P:E.classical}, becomes 
\begin{equation}
\label{P:L2f.3}
	\mathcal{L}_{2F} = i \bar{\theta} \Pi_+ \gamma_\alpha \unih^{\alpha\beta}  \mathcal{D}_\beta \theta~. 
\end{equation}
Formally, we can now apply the manipulations described in section~\ref{NambuGoto}, albeit there may be ambiguities depending on which components are present in the Ramond-Ramond fields. 

We will stop here as far as the general case is concerned and turn our attention, in the next section, to applications in which the actions $\mathcal{L}_{2B}$ and $\mathcal{L}_{2F}$ are free of ambiguities. 
 
\section{Hagedorn temperature}
\label{Hage}

In this section, we will reformulate the worldsheet calculation of the Hagedorn temperature in confining gauge theories that was presented in \cite{Bigazzi:2023oqm, Bigazzi:2024biz}. The formalism will be based on the non-standard semiclassical treatment described in subsection~\ref{P:nonstandard}. 

Consider a class of simple setups often called a ``string at the tip of a cigar''. It contains a classical string worldsheet extending along the (non-compact) time direction and winding once around a compact spacelike cycle, whose circumference measures the inverse temperature. Moreover, the worldsheet is fixed to a special location (the tip of the cigar) by the requirement of a vanishing extrinsic curvature for the classical worldsheet. For simplicity, we assume a 10-d background without Kalb-Ramond field. The geometry of such a configuration is trivial. The induced metric is flat and all extrinsic curvature components as well as the normal bundle connections vanish,
\begin{align}
\label{Hage:trivial.geom}
	\Gamma^\alpha{}_{\beta\gamma} = 0~, \quad K^i{}_{\alpha\beta} =0~, \quad A_{ij\alpha}=0~.
\end{align}
Moreover, the pull-back of the spacetime Riemann tensor is such that the bosonic sector of the fluctuation fields is described by eight massive scalar fields, each with an action of the form [this derives from \eqref{P:L2B.1}, while the contribution \eqref{P:L2B.2} vanishes]
\begin{align}
\notag
	S_{2B} &= -\frac1{4\pi\alpha'} \int \rmd^2 \sigma \unih^{\alpha\beta} \left[(\partial_\alpha\chi)(\partial_\beta \chi)
	+\frac12 m_b^2 g_{\alpha\beta} \chi^2 \right]\\
\label{Hage:action.scalars}
	&= -\frac1{4\pi\alpha'} \int \rmd^2 \sigma \left[ \unih^{\alpha\beta} (\partial_\alpha\chi)(\partial_\beta \chi)
		+ \mu_b^2 \chi^2 \right]~,
\end{align}
where we have abbreviated
\begin{equation}
\label{Hage:mu.b}
	\mu_b^2= \sqrt{-g}\Omega m_b^2~.
\end{equation}
We have explicitly written the first line in \eqref{Hage:action.scalars}, because it is the combination $m_b^2 g_{\alpha\beta}$ that arises in \eqref{P:L2B.1} from the geometrical properties of the background worldsheet.  
The auxiliary metric is conformal to the full string metric $G_{\alpha\beta}$,
\begin{equation}
\label{Hage:aux.metric}
	\sqrt{-G} G^{\alpha\beta} = \unih^{\alpha\beta}
\end{equation}
by virtue of its field equation. The string metric $G_{\alpha\beta}$, in turn, is to be distinguished from the background induced metric, $g_{\alpha\beta}$. In fact, one has 
\begin{equation}
\label{Hage:string.metric}
	G_{\alpha\beta} = g_{\alpha\beta} + t_{\alpha\beta}~, 
\end{equation}
where $t_{\alpha\beta}$ is a contribution from the quantum fluctuations, which needs to be calculated.

Similarly, the fermionic sector of the fluctuations is described by the action \eqref{P:L2f.3}, which, for this class of cases, reads\footnote{A conformal rescaling may be needed to obtain \eqref{Hage:action.fermions} from \eqref{P:L2f.3}.} 
\begin{equation}
\label{Hage:action.fermions}
	S_{2F} = -\frac{i}{2\pi\alpha'} \int \rmd^2 \sigma \bar{\psi} \left[ \unih^{\alpha\beta} \gamma_\alpha \nabla_\beta + \mu_f  \right]\psi~,
\end{equation}
where we have introduced 
\begin{equation} 
\label{Hage:mu.f}
	\mu_f = \left(\sqrt{-g}\Omega\right)^\frac12 m_f~.
\end{equation}

the gamma matrices satisfy $\gamma_{(\alpha} \gamma_{\beta)} = \unih_{\alpha\beta}$.
The constant values of $m_f$ arise as the eigenvalues of the matrix $\tilde{S}$ of \eqref{NG:S.tilde}. One has ``mass matching'' between the bosons and fermions, 
\begin{equation}
\label{Hage:mass.match}
	\sum\limits_b m_b^2 = \sum\limits_f m_f^2~, \qquad \sum\limits_b \mu_b^2 = \sum\limits_f \mu_f^2~,
\end{equation}
which ensures the finiteness of the one-loop partition function in the Nambu-Goto formalism, as discussed in general in section~\ref{NG:anomaly}.

The system is completed by identifying the quantum contribution $t_{\alpha\beta}$ in \eqref{Hage:string.metric} as\footnote{This is done by comparison with the zeroth order action. The apparent mismatch by a factor of $\frac12$ in the fermionic term with respect to \eqref{Hage:action.fermions} is explained by the fact that the string metric gets a fermionic contribution only from the metric term with $\unih^{\alpha\beta}$ in the Polyakov action, whereas the action \eqref{Hage:action.fermions} receives an (equal) contribution  also from the Wess-Zumino term.}
\begin{align}
\label{Hage:t.def}
	t_{\alpha\beta} &= \sum\limits_b \vev{(\partial_\alpha\chi)(\partial_\beta \chi)
		+\frac12  \Omega^{-1} \unig_{\alpha\beta} \mu_b^2 \chi^2}\\
\notag &\quad 
		+ \sum\limits_f \vev{\bar{\psi} \left[\gamma_{(\alpha} \nabla_{\beta)} +\frac12  \Omega^{-1}\unig_{\alpha\beta} \mu_f \right]\psi}~.
\end{align}
Assuming that the background value of the auxiliary metric $\unih^{\alpha\beta}$ is constant, which is reasonable for a flat induced metric, the vacuum expectation values in \eqref{Hage:t.def} can be evaluated using canonical quantization. 
To keep the presentation as general as possible, we parameterize $\unih^{\alpha\beta}$ in the form 
\begin{equation} 
\label{Hage:h.param}
	\unih^{\alpha\beta} = \frac1{H} \begin{pmatrix}
	-1+h_1 & h_2 \\
	h_2 & 1+h_1 
	\end{pmatrix}~, \qquad H = \sqrt{1-h_1^2+h_2^2}~,
\end{equation}
assuming that the combination under the square root in $H$ is positive.
The details of the canonical quantization are deferred to appendix~\ref{quant}.
The result for $t_{\alpha\beta}$, summing all the contributions from \eqref{quant:t.scalar.result} and \eqref{quant:t.fermion.result}, is 
\begin{equation}
\label{Hage:t.result}
	t_{\alpha\beta} = -\alpha' \left[ \left( \Delta - \tilde{\Delta} \right) T_{\alpha\beta} 
	+\frac{H}{1-h_1} \tilde{\Delta} \left( \Omega^{-1} \unig_{\alpha\beta} - \unih_{\alpha\beta}\right)\right]~,	
\end{equation}
with 
\begin{align}
\label{Hage:Delta}
	\Delta &= \sum\limits_b \Delta_b - \sum\limits_f \Delta_f~,\\
\label{Hage:tilde.Delta}
	\tilde{\Delta} &= \sum\limits_b \tilde{m}_b^2 \Delta'_b - \sum\limits_f \tilde{m}_f^2 \Delta'_f~.
\end{align}
We refer to appendix~\ref{quant} for the definitions, in particular \eqref{quant:delta.b} and \eqref{quant:delta.f} for $\Delta_{b,f}$, \eqref{quant:delta.prime.b} and \eqref{quant:delta.prime.f} for $\Delta'_{b,f}$, \eqref{quant:lambda.n} and \eqref{quant:lambda.r} for $\tilde{m}_{b,f}^2$, as well as \eqref{quant:T.ab.def} for $T_{\alpha\beta}$. We remark that the mass matching \eqref{Hage:mass.match} also holds for $\tilde{m}_{b,f}^2$. The quantity $\Delta_b$ ($\Delta_f$) corresponds to minus one half the zero point or Casimir energy for a two-dimensional massive bosonic (fermionic) field, with twisted boundary conditions.

The tensor $t_{\alpha\beta}$ is traceless, $\unih^{\alpha\beta} t_{\alpha\beta}=0$. Hence, combining \eqref{Hage:string.metric} and \eqref{Hage:aux.metric} shows that 
\begin{equation}
\label{Hage:sqrt.G}
	\sqrt{-G} = \sqrt{-g}\Omega~.
\end{equation}

$\Delta$ and $\tilde{\Delta}$ can be evaluated using zeta function regularization of the infinite sums (see appendix~\ref{quant}), which results in
\begin{align}
\label{Hage:Delta.series}
	\Delta &= 1 -\frac12 \sum\limits_b \tilde{m}_b + \ln 2  \sum\limits_b \tilde{m}_b^2 +\mathcal{O}(\tilde{m}_{b,f}^4)~,\\
\label{Hage:tilde.Delta.series}
	\tilde{\Delta} &= -\frac14 \sum\limits_b \tilde{m}_b + \ln 2  \sum\limits_b \tilde{m}_b^2 +\mathcal{O}(\tilde{m}_{b,f}^4)~.
\end{align}
Here, the mass matching has been used, which is crucial the render the terms quadratic in the masses finite. It is interesting to note that the mass matching would not be needed, if one only wanted to evaluate the term with $T_{\alpha\beta}$ in \eqref{Hage:t.result}, because the potentially divergent term exactly cancels between $\Delta$ and $\tilde{\Delta}$. However, it is needed for the other term in \eqref{Hage:t.result}.

It remains to substitute \eqref{Hage:t.result} into \eqref{Hage:string.metric}, which leads to
\begin{equation}
\label{Hage:mat.eq}
	\left(\Delta - \tilde{\Delta}\right) T_{\alpha\beta} + \left[\frac{H\tilde{\Delta}}{(1-h_1)} - \frac{\Omega\sqrt{-g}}{\alpha'} \right] \left(\Omega^{-1}\unig_{\alpha\beta}-\unih_{\alpha\beta} \right) =0~, 
\end{equation}
where also \eqref{Hage:aux.metric} and \eqref{Hage:sqrt.G} have been used. From this matrix equation, which is equivalent to two equations, because the matrix is symmetric and traceless, one can find the two unknown parameters in the problem. One has several options how to proceed. We will present two calculations, which obviously lead to equivalent results.

The first option, which appears natural from the point of view of the classical background worldsheet, is to start with coordinates, in which the background induced metric is very simple. Let us take $\unig_{\alpha\beta}=\eta_{\alpha\beta}$, which implies $\sqrt{-g}=g_{\sigma\sigma}$. $g_{\sigma\sigma}$ is fixed by the inverse temperature $\beta$ (in supergravity units),
\begin{equation}
\label{Hage:gss}
	g_{\sigma\sigma} = \left(\frac{\beta}{2\pi}\right)^2~. 
\end{equation}
The two unknown parameters, which are to be determined by solving \eqref{Hage:mat.eq}, are $h_1$ and $h_2$ in the parameterization \eqref{Hage:h.param}.
The matrix $T_{\alpha\beta}$ is explicitly given in \eqref{quant:T.ab.def}. It is easy to see that the solution of \eqref{Hage:mat.eq} must have $h_2=0$, so that \eqref{Hage:mat.eq} reduces to the single equation
\begin{equation}
\label{Hage:eq.h1}
	\Delta - (1-h_1) \tilde{\Delta} - \frac{g_{\sigma\sigma} h_1}{(1+h_1)\alpha'} =0~.
\end{equation} 
Because $\Delta$ and $\tilde{\Delta}$ contain $h_1$ hidden in the mass parameters,\footnote{Equation \eqref{Hage:m.tilde} follows from \eqref{Hage:mu.b} and \eqref{quant:lambda.n}, together with $\Omega = 1/H$ and $h_2=0$.}  
\begin{equation}
\label{Hage:m.tilde}
	\tilde{m}_{b,f}^2 = \frac{g_{\sigma\sigma}}{1+h_1}m_{b,f}^2 ~,
\end{equation} 
\eqref{Hage:eq.h1} is an implicit equation for the parameter $h_1$ as a function of the inverse temperature $\beta$.

Following the argument by Atick and Witten \cite{Atick:1988si}, the Hagedorn temperature can be identified as the temperature at which the auxiliary metric degenerates, \ie for $h_1=1$.\footnote{Atick and Witten \cite{Atick:1988si} state that, as one approaches the Hagedorn temperature, a string wrapping once around the compact direction becomes massless. The additional massless (long-range) degree of freedom is a telltale sign of the phase transition. A massless string has a null worldsheet, which has a degenerate metric.}
Hence, it can be found by the relations
\begin{equation}
\label{Hage:Hagedorn.temp}
	\Delta = \frac{\beta_H^2}{8\pi^2\alpha'}~, \qquad \tilde{m}_{b,f}^2 = \frac{\beta_H^2}{8\pi^2}m_{b,f}^2~,
\end{equation}  
reproducing the results of \cite{Bigazzi:2023oqm, Bigazzi:2024biz}. 

The second option is to use coordinates such that $\unih^{\alpha\beta}=\eta^{\alpha\beta}$, so that the canonical quantization is simpler. In fact, one has $h_1=h_2=0$, $H=1$ in \eqref{Hage:h.param}, so that $T_{\alpha\beta}=\mathrm{diag}(1,1)$. In this case, the two metric components $g_{\tau\tau}$ and $g_{\tau\sigma}$ are the unknown parameters, $g_{\sigma\sigma}$ being fixed by the inverse temperature as in \eqref{Hage:gss}. It is again easy to see that one needs $g_{\tau\sigma}=0$, so that \eqref{Hage:mat.eq} reduces to the single equation 
\begin{equation}
\label{Hage:eq.g}
	\Delta + \frac{2g_{\tau\tau}}{g_{\sigma\sigma}-g_{\tau\tau}} \tilde{\Delta} - \frac{g_{\sigma\sigma}+g_{\tau\tau}}{2\alpha'} =0~,
\end{equation}
where the mass parameters in $\Delta$ and $\tilde{\Delta}$ are 
\begin{equation}
\label{Hage:m.tilde2}
	\tilde{m}_{b,f}^2 = \frac12 \left(g_{\sigma\sigma}-g_{\tau\tau}\right) m_{b,f}^2~.
\end{equation}
These two equations implicitly determine $g_{\tau\tau}$ as a function of the inverse temperature. As above, the worldsheet degenerates at the Hagedorn temperature, \ie for $g_{\tau\tau}=0$, so that one again finds \eqref{Hage:Hagedorn.temp}.

It is also worth noting that at very low temperatures, \ie for very large $g_{\sigma\sigma}$, the mass parameters $\tilde{m}_{b,f}$ are very large and both $\Delta$ and $\tilde{\Delta}$ are exponentially small, as one can see via more appropriate integral representations. Therefore, \eqref{Hage:eq.h1} and \eqref{Hage:eq.g} indicate that the background induced metric and the auxiliary metric become essentially conformal to each other, which confirms that the semiclassical analysis in the Nambu-Goto formulation is sufficient at low temperatures. 
\section{Examples}
\label{Examples}

In this section, we will provide a number of examples in which the simplifying assumptions of section~\ref{Hage} hold. This means that they (or most of them) show a ``cigar'' submanifold with a vanishing $B$ field. The final result for flat space is exact and we will show how our formalism naturally leads to the well-known value reported in the literature \cite{Kogan:1987jd, Atick:1988si}. In all the other scenarios we will go as far as the next-to-next-to-leading order (NNLO) in the holographic limit.  Let us stress that the NNLO term provided here is not complete. Indeed, it does not include terms that would be captured by a quartic analysis of the worldsheet sigma model \cite{Bigazzi:2024biz} or by the effective approach \cite{maldanotes,Urbach:2022xzw,Urbach:2023npi,Ekhammar:2023glu,Bigazzi:2023hxt,Ekhammar:2023cuj,Harmark:2024ioq, Horowitz:1997jc, Bigazzi:2024biz}. In any case, the general structure of our predictions looks pretty much universal, in agreement with \cite{Bigazzi:2024biz}. Moreover, the results in subsections \ref{Ex:TwistedAdS} and \ref{Ex:TwistedKW} provide new predictions for the Hagedorn temperature for the twisted compactification of $\mathcal{N}= 4$ 4D Super Yang-Mills and Klebanov-Witten CFT, respectively. 

Remarkably, in global $AdS$ setups, the analysis of the Hagedorn temperature has been extended up to NNNLO  \cite{Ekhammar:2023glu,Bigazzi:2023hxt,Ekhammar:2023cuj}. In particular, in the specific examples of $AdS_5$ and $AdS_4$, respectively dual to $N = 4$ Super Yang-Mills on $S^3$ and to the ABJM theory on $S^2$, the holographic results, based on an effective approach, for the Hagedorn temperature to NNNLO match those obtained via Quantum Spectral Curve (QCS) integrability methods \cite{Ekhammar:2023glu, Ekhammar:2023cuj, Harmark:2017yrv, Harmark:2018red, Harmark:2021qma}. Nevertheless, developing a string worldsheet-based approach to study higher order corrections to Hagedorn temperature in global $AdS$ backgrounds remains challenging due to the difficulty in finding a regular, non-degenerate classical string configuration around which to carry out the semiclassical analysis of the fluctuations. Hence, for these reasons, in this work we cannot include a study of the Hagedorn temperature for global $AdS$ solutions and their dual QFTs.

\subsection{Flat spacetime}
\label{Ex:flat}

As the first example, we consider flat spacetime with one compact dimension of size $\beta$ and a string worldsheet that wraps once around this compact dimension and otherwise extends along the time direction. This, of course, will reproduce the superstring Hagedorn temperature calculated long ago \cite{Kogan:1987jd, Atick:1988si}. All bosonic and fermionic fluctuations are massless. Setting $\tilde m_{b\,,f} =0$ in \eqref{Hage:Delta.series} and \eqref{Hage:tilde.Delta.series}, we immediately have  $\Delta=1$ and $\tilde{\Delta}=0$.
So, let $x^1\sim x^1+\beta$ be the compact direction. Without loss of generality, we consider a classical string worldsheet parameterized by $\tau\in(-\infty,\infty)$, $\sigma \in[0,2\pi)$, with the embedding functions
\begin{equation}
\label{Ex:flat.embed}
	x^0= \frac{\beta}{2\pi} \tau~,\qquad x^1= \frac{\beta}{2\pi} \sigma~,
\end{equation}
so that the induced metric is 
\begin{equation}
\label{Ex:flat.g.induced}
	g_{\alpha\beta} = \left(\frac{\beta}{2\pi}\right)^2 \eta_{\alpha\beta}~.
\end{equation}
For a general inverse temperature $\beta$, the solution of the auxiliary metric parameter $h_1$ is found by solving the implicit equation in \eqref{Hage:eq.h1} and reads
\begin{equation}
\label{Ex:flat.h1}
	h_1 = \frac1{\frac{\beta^2}{4\pi^2 \alpha'}-1}~.
\end{equation}
The critical temperature at which the worldsheet degenerates, \ie the superstring Hagedorn temperature, can be found by setting $h_1=1$ in \eqref{Ex:flat.h1}, 
\begin{equation}
\label{Ex:flat.TH}
	T_H = \frac1{\beta_H} = \frac1{\sqrt{8\alpha'}\pi}~.
\end{equation}
In alternative to \eqref{Ex:flat.embed}, one may use $\unih^{\alpha\beta}=\eta^{\alpha\beta}$ and parameterize the classical embedding by 
\begin{equation}
\label{Ex:flat.embed2}
	x^0= M \tau~,\qquad x^1= \frac{\beta}{2\pi} \sigma~,
\end{equation}
implying an induced metric with parameters
\begin{equation}
\label{Ex:flat.induced2}
	g_{\tau\tau} = -M^2~, \qquad 
	g_{\sigma\sigma} =\frac{\beta^2}{4\pi^2}~,\qquad g_{\tau\sigma}=0~.
\end{equation}
Then, from \eqref{Hage:eq.g} one finds
\begin{equation}
\label{Ex:flat.Msol}
	M^2 = \frac{\beta^2}{4\pi^2} - 2 \alpha'~.
\end{equation}
Setting $M=0$ yields again the Hagedorn temperature \eqref{Ex:flat.TH}.\footnote{The parameter $M$ is interpreted as the rest mass of the classical string. Thence derives the statement that the wound string becomes massless at the Hagedorn temperature \cite{Atick:1988si}.}

\subsection{Witten-Yang-Mills}
\label{Ex:WYM}

The Witten-Yang-Mills (WYM) background \cite{Witten:1998zw} was described in detail in \cite{Bigazzi:2022gal, Bigazzi:2023oqm}. We refer to \cite{Bigazzi:2023oqm} for the notation of the description of the background string worldsheet.
The classical string worldsheet bound to the tip of the cigar and wrapping a compactified direction possesses scalar fluctuations whose mass matrix is determined by the relation
\begin{equation}
\label{Ex:R.special}
	R_{\alpha i \beta j} = -\frac12  m_i^2 \delta_{ij}\, g_{\alpha\beta}~.
\end{equation}
In particular, in the present case, there are six massless and two massive scalar fields with
\begin{equation}
\label{Ex:WYM.masses}
m^2 = \frac{9}{4m_0R^3}~. 
\end{equation}
Let us work in the approach with an embedding similar to the one in \eqref{Ex:flat.embed}. Then, we have
\begin{equation}
\label{Ex:WYM.sqrt.g}
	\sqrt{-g} = m_0R^3 \rho^2~, 
\end{equation}
where $\rho$ is the coordinate radius of the compactified dimension. Given $\rho$, the auxiliary metric parameter $h_1$ is determined by \eqref{Hage:eq.h1}, where $\Delta$ and $\tilde{\Delta}$ are functions of 
\begin{equation}
\label{Ex:WYM.mass.tilde}
	\tilde{m}^2=\frac{\sqrt{-g}\,\eta_{\sigma\sigma}}{(1+h_1)}m^2=\frac{9\rho^2}{4(1+h_1)}~,
\end{equation}
see \eqref{Hage:m.tilde}, so that the solution for $h_1$ is not explicit.
In order to find the critical radius $\rho_H$, we set $h_1=1$ and solve \eqref{Hage:Hagedorn.temp}, which becomes
\begin{equation}
\label{Ex:WYM.rho_H.eq}
	1 -\frac{3\rho_H}{2\sqrt{2}} + \frac94 \ln 2\, \rho_H^2 +\cdots = \frac{m_0R^3\rho_H^2}{2\alpha'} 
	= \frac{\lambda}{12} \rho_H^2 = \pi T_s \rho_H^2~,
\end{equation}
where we have expressed the right hand side alternatively in terms of the `t~Hooft coupling $\lambda$ or the dimensionless string tension $T_s$. The solution is
\begin{equation}
\label{Ex:WYM.rho_H}
	\rho_H = \frac1{\sqrt{\pi T_s}} \left[ 1 - \frac{3}{4\sqrt{2\pi T_s}} 
	+ \frac{1+8\ln 2}{2} \left(\frac{3}{4\sqrt{2\pi T_s}}\right)^2 +\cdots\right]~.
\end{equation}

The leading-order (LO) term is just the inverse superstring Hagedorn temperature \eqref{Ex:flat.TH} in different units. The 
next-to-leading-order (NLO) term displays the correct coefficient when compared with the results coming from the effective approach \cite{Urbach:2023npi}. Moreover, the part with $\ln 2$ in the next-to-next-to-leading-order (NNLO) term has the correct coefficient that is needed as an input in the effective model approach \cite{Bigazzi:2024biz}. The rest of the NNLO term will be modified at higher orders due to string interactions, which is well captured within the effective model \cite{Bigazzi:2024biz}.

\subsection{Type IIB supergravity on wrapped D5-branes}
\label{Ex:Nunez}

The next example is the $4+1$ gauge theory whose dual was presented in section~2 of\cite{Nunez:2023xgl}, which consists of a confining type IIB supergravity solution deriving from a wrapped D5-brane setup. 
This solution can be written as follows,
\begin{align}
\notag
	\rmd s^2 &= m_0 R^3 r \left\{ \rmd x_{1,4}^2 + \frac{f(r)}{(1+q^2)^2} \rmd \varphi^2 + \frac{1}{r^2 f(r)}  \rmd r^2 
	+ \left[ \omega_1^2 +\omega_2^2 + \left( \omega_3 - \frac{q}{1+q^2} \zeta(r) \rmd \varphi\right)^2 \right] \right\}~,\\
\notag
	F_3 &= R^2 \left[2 \Omega_3 + \frac{q}{1+q^2} \rmd \left( \zeta(r) \omega_3 \wedge \rmd \varphi \right)\right]~,\\
\label{Ex:Nunez.bg}
	\e{\Phi} &= m_0 R r~,
\end{align}
where the functions $f(r)$ and $\zeta(r)$ are given by 
\begin{align}
	f(r) &= \frac{(r^2-1)(r^2+q^2)}{r^4}~,\\
	\zeta(r) &= 1- \frac1{r^2}~, 
\end{align}
and the left-invariant one-forms $\omega_1$, $\omega_2$, $\omega_3$ and the volume 3-form $\Omega_3$ are defined in terms of three angular variables $\left[\theta, \phi,\psi\right]$,\footnote{Note that our $\omega_1$, $\omega_2$, $\omega_3$ differ from those of \cite{Nunez:2023xgl} by a factor of two.}
\begin{subequations}
\label{oneforms}
\begin{align} 
	\omega_1 &= \frac12 \left( \cos\psi\rmd\theta + \sin\psi\sin\theta \rmd\phi\right)~,\\
	\omega_2 &= \frac12 \left(-\sin\psi \rmd\theta + \cos\psi\sin\theta\rmd\phi\right)~,\\
	\omega_3 &= \frac12 \left(\rmd\psi + \cos\theta\rmd\phi\right)~,\\
	\Omega_3 &= \omega_1 \wedge \omega_2 \wedge \omega_3 = \frac18 \sin \theta \rmd \theta \wedge \rmd \phi \wedge \rmd \psi~.
\end{align}
\end{subequations}
Our parameters $m_0$, $R$ and $q$ are related to the parameters of \cite{Nunez:2023xgl}, after reintroducing $g_s$ and $\alpha'$, by\footnote{$r_-^2=-q^2 r_+^2$ is negative in \cite{Nunez:2023xgl}, which is awkward to work with.} 
\begin{equation}
\label{Ex:Nunez.R.def}
	R^2 = g_s \alpha' N~,\qquad m_0 R = \frac{r_+}{g_s}~,  \qquad q = -\frac{\sqrt{2}Q}{r_+^2}~,
\end{equation}
and we have rescaled some coordinates of the original background in \eqref{Ex:Nunez.bg}: $r \to r_+r$, so that now $r\in[1,\infty)$; $\varphi\to\frac{L_\varphi}{2\pi} \varphi$, so that now $\varphi\in[0,2\pi)$; $x_i\to \sqrt{N} x_i$.

The background worldsheet at the tip of the cigar is given by the embedding 
\begin{align}
x^0 = \rho \tau\,,\quad x^1 = \rho \sigma\,,\quad r=1\,.
\end{align}
The latter leads to an induced metric of the form
\begin{align}
g_{\alpha\beta} = R^3m_0\rho^2 \eta_{\alpha\beta}\,.
\end{align}
As in the WYM case, here we can use a constant $r>1$ to avoid coordinate singularities and then take the $r\to1$ limit. Remarkably, we find that the worldsheet is geometrically trivial at $r=1$, \ie the conditions in (\ref{Hage:trivial.geom}) are satisfied.

Then, there are again six massless and two massive scalars. For the latter, we have that
\begin{equation}
R_{\alpha i \beta j} = -\frac{1+q^2}{2m_0 R^3}g_{\alpha \beta}\delta_{ij} = -\frac{1}{2}\rho^2 (1+q^2) \eta_{\alpha\beta}\delta_{ij}\,,
\end{equation}
from which we can identify the masses $m_i^2 = m^2$\, that read
\begin{equation}
m^2 = \frac{1+q^2}{m_0 R^3}\,.
\end{equation}
Moreover, using \eqref{Hage:m.tilde}, we have that 
\begin{equation}
\tilde m^2 = \frac{1+q^2}{(1+h_1)}\rho^2\,.
\end{equation}
The critical dimensionless radius that follows from \eqref{Hage:Hagedorn.temp} is then given by
\begin{equation}
\rho_H = \frac{1}{\sqrt{\pi T_s}}\left[1-\frac{\sqrt{1+q^2}}{2}\frac{1}{\sqrt{2\pi T_s}}+ \frac{1+8\ln2}{2}\left(\frac{\sqrt{1+q^2}}{2}\frac{1}{\sqrt{2\pi T_s}}\right)^2+\cdots\right]\,.
\end{equation}
It coincides with the expression \eqref{Ex:WYM.rho_H} for the WYM case, if we modify the coefficients of the $1/{\sqrt{\pi T_s}}$ terms by the shift $3/4 \to \sqrt{1+q^2}/2$.

\subsection{\texorpdfstring{$AdS_5\times S^5$}{AdS5xS5} soliton}
\label{Ex:TwistedAdS}

In this example we aim to analyze the Hagedorn temperature for a string in the twisted compactified soliton $AdS_5\times S^5$ background presented in \cite{Anabalon:2021tua} and analysed later in, \eg \cite{Kumar:2024pcz,Chatzis:2024top,Chatzis:2024kdu,Castellani:2024ial}. In particular, this is a type IIB supergravity solution derived from the $AdS_5\times S^5$ solution by compactifying along one Poicar\'e direction of $AdS$. This procedure in general breaks SUSY, because the scalars and gauge fields, and the fermions are assigned periodic and anti-periodic boundary conditions, respectively. Interestingly, instead, some SUSY (in particular four supercharges) can be preserved by turning on a background gauge field $\mathcal{A}$ mixing a $U(1)$ of R-symmetry with the $U(1)$ isometry associated with the compact circle \cite{Anabalon:2021tua,Kumar:2024pcz}. The metric of the supergravity solution then reads
\begin{align}
\notag
  \rmd s^2 _{10} &= \frac{r^2}{\ell^2} (-\rmd t^2+\rmd x_1^2 + \rmd x_2^2 +  f(r)\rmd \phi^2) + \frac{\ell^2\rmd r^2}{ r^2 f(r)}+\ell^2 \sum_{i=1}^3 \left[\rmd  \mu_i^2 + \mu_i^2 \left( \rmd \phi_i + \mathcal{A} \right)^2\right]~,\\
\label{Ex:metric-ARxS5}
  f(r) &= 1-\frac{Q^6\ell^{12}}{r^6}\,,\quad \mathcal{A} =  \ell^4 Q^3\left( \frac{1}{r^2}- \frac{1}{Q^2\ell^4}\right)\rmd \phi\,,\quad \ell^4 = 4\pi g_s  \alpha^{\prime\,2}N\,.
\end{align} 
Here, for a lighter notation, we have defined 
\begin{equation}
\mu_1= \sin\theta \sin\varphi,~\mu_2= \sin\theta \cos\varphi, ~\mu_3= \cos\theta\,.
\end{equation}
Moreover, the end of the spacetime (``tip of the cigar'') is given by the $r=r_0 =Q$, for which $f(r_0)=0$, while the domain of $\phi$ is fixed to $[0, \frac{2\pi}{3Q}]$ by demanding the absence of a conical singularities at the tip. Finally the RR-form content of the solution is given by
\begin{align}
  F_5 &= ( 1+ *) G_5\,, \qquad 
  G_5= -4 \, \mathrm{vol}_{5}  -2Q^3 \ell^4 J_2\wedge\rmd t\wedge \rmd x_1 \wedge \rmd x_2\,, \\
\label{Ex:RR-S5}
  J_2 &=  \sum_{i=1}^3 \mu_i \rmd \mu_i \wedge \left(\rmd \phi_i +\mathcal{A}\right)\,.
\end{align}
Is useful to convert to dimensionless coordinates by letting $x_i\to \ell x_i$ and $r\to Q\ell^2 r$, where the latter implies  $r_0=1$. Moreover, it is convenient to further rescale the angle $\phi$ by
\begin{eqnarray}
    \label{Ex:period}
	\phi\to \frac{1}{3Q\ell}\phi\,,\quad \phi \sim \phi +2\pi\,.
\end{eqnarray}
In these dimensionless coordinates, \eqref{Ex:metric-ARxS5} reads
\begin{align}
\notag
  \rmd s^2_{10} &= \ell^4 Q^2 r^2 \left(-\rmd t^2+\rmd x_1^2 + \rmd x_2^2 + \frac{f(r)}{9Q^2\ell^2} \rmd \phi^2 \right)+ \frac{\ell^2\rmd r^2}{ r^2 f(r)}+ \ell^2\sum_{i=1}^3 \left[\rmd \mu_i^2 + \mu_i^2 \left( \rmd \phi_i + \mathcal{A} \right) ^2\right]\,,\\
\label{Ex:metric-ARxS5_dimensionless}
  f(r) &= 1-\frac{1}{r^6}\,,\quad \mathcal{A} =   -\frac{1}{3}\left( 1-\frac{1}{r^2}\right)\rmd \phi\,,\quad \ell^4 = 4\pi g_s \alpha^{\prime\,2} N\,,
\end{align} 
while the five-form \eqref{Ex:RR-S5} becomes
\begin{align}
  F_5 &= ( 1+ *) G_5\,, \qquad 
  G_5= -4 Q\ell^6\, \mathrm{vol}_{AdS_5}  -2Q^3 \ell^7 J_2\wedge\rmd t\wedge \rmd x_1 \wedge \rmd x_2\,, \\
\label{Ex:RR-S5.dimensionless}
  J_2&=  \sum_{i=1}^3 \mu_i \rmd \mu_i \wedge \left(\rmd \phi_i +\mathcal{A}\right)\,.
\end{align}
Similarly to the ansatz in subsection \ref{Ex:WYM} in the case of WYM we can consider the background string embedding \footnote{Here, we have compactified the $x_1$ direction such that $x_1 \sim x_1 + \beta$.}
\begin{eqnarray}
\label{Ex:AdS5_string}
&&t = \rho \tau\,,\quad x_1 =  \rho \sigma\,,\quad \rho \equiv
\frac{\beta}{2\pi}\,.
\end{eqnarray}
Moreover, the string is required to live at the minimum radius surface, namely at $r=1$, so that the background induced metric is 
\begin{equation}
    g_{\alpha\beta} = Q^2\ell^4 \rho^2 \eta_{\alpha\beta}\,.
\end{equation}
This background worldsheet is again geometrically trivial, with 
\begin{equation}
    \Gamma^\alpha{}_{\beta\gamma} = 0\,,\quad K^i{}_{\alpha\beta} = 0\,,\quad A_{ij\alpha} =0\,.
\end{equation}
Then, following similar steps as in the previous examples, we can calculate the masses of the bosonic modes. There are again six massless and two massive modes, the latter determined by 
\begin{equation}
    R_{\alpha i \beta j} =-\frac{1}{2}m^2 \delta_{ij}g_{\alpha\beta}\,, \quad m^2 = \frac{6}{Q^2\ell^4}\,,
\end{equation}
for the two components extending along the cigar ($r$ and $\phi$). According to \eqref{Hage:m.tilde}, the effective mass parameter for these modes is
\begin{equation}
    \tilde m^2 = \frac{\sqrt{-g}\,\eta_{\sigma\sigma}}{1+h_1} m^2 = \frac{6\rho^2}{1+h_1}\,.
\end{equation}
In order to extract the Hagedorn radius $\rho_H$, we set $h_1= 1$ and solve \eqref{Hage:Hagedorn.temp}, which becomes 
\begin{equation}
\label{Ex:AdS5.rho_H.eq}
	1 -\sqrt{3}\rho_H + 6\ln 2\, \rho_H^2 +\cdots = \frac{Q^2 \ell^4 \rho_H^2}{2\alpha'} 
	=  \pi T_s \rho_H^2~,
\end{equation}
where the dimensionless string tension is  
\begin{equation}
\label{Ex:AdS5.Ts}
	2\pi \alpha^\prime T_s = Q^2\ell^4\,.
\end{equation}
This finally leads to 
\begin{equation}
\label{Ex:AdS.rho_H}
	\rho_H = \frac1{\sqrt{\pi T_s}} \left[ 1 - \sqrt{\frac{3}{2}}\frac{1}{\sqrt{2\pi T_s}} 
	+ \frac{1+8\ln 2}{2} \left(\sqrt{\frac{3}{2}}\frac{1}{\sqrt{2\pi T_s}} \right)^2 +\cdots\right]~,
\end{equation}
which again has a form similar to the WYM Hagedorn temperature \eqref{Ex:WYM.rho_H} upon replacing $3/4 \to \sqrt{3/2}$.

\subsection{Twisted compactification of Klebanov-Witten}
\label{Ex:TwistedKW}

As a last example, let us consider a type-IIB supergravity solution presented in, \eg \cite{Chatzis:2024kdu,Chatzis:2024top,Castellani:2024pmx} arising from a twisted compactification of the $\mathcal{N}=1$ Klebanov-Witten background. In particular, the latter in given by\footnote{Contrary to \cite{Chatzis:2024kdu,Chatzis:2024top}, here we directly introduce the solution with dimensionless coordinates.}
\begin{align}
\label{Ex:AdS5xT11}
	\rmd s^2_{10}&= \ell^4 Q^2 r^2 \left(-\rmd t^2+\rmd x_1^2 + \rmd x_2^2 + \frac{1}{9Q^2\ell^2} f(r)\rmd \phi^2 \right)+\ell^2 \frac{\rmd r^2}{ r^2 f(r)} \\
\notag &\quad +\ell^2\left[\frac{1}{6}\sum_{i=1}^2\left(\rmd \theta_i^2+\sin^2\theta_i\rmd \phi_i^2\right) + \frac{1}{9}\left(\rmd \psi + \sum_{i=1}^2\cos\theta_i\rmd \phi_i+3\mathcal{A}\right)^2\right]\,,
\end{align}
with the warp factor and fibration gauge field defined as in the $AdS_5\times S^5$ soliton \eqref{Ex:metric-ARxS5}, while
\begin{equation}
    \ell^4 = \frac{27}{4}\pi g_s \alpha^{\prime\,2} N\,.
\end{equation}
The shrinking circle $\phi$ has period $2\pi$ as in equation \eqref{Ex:period}, and the tip of the cigar is placed at $r=1$. The solution is completed by the following Ramond-Ramond fields
\begin{align}
\notag 
  F_5 &=\left(1+\ast\right) \,G_5\,, \qquad G_5=-4Q \,\ell^6 \text{vol}_{AdS_5} -2Q^3 \ell^7\,J_2\wedge\rmd t\wedge \rmd x_1 \wedge \rmd x_2\,,\\
\label{AdS5xT11_G5}
  J_2&=-\frac{1}{6}\left(\rmd \theta_1 \wedge \sin\theta_1\rmd \phi_1 +\rmd \theta_2 \wedge \sin\theta_2\rmd \phi_2\right)\,.
\end{align}
Remarkably, this type-IIB supergravity background is confining and preserves four supercharges. 

As in the previous examples, the background string worldsheet 
\begin{equation}
\label{Ex:AdS5xT11_string}
	t = \rho \tau\,,\quad x_1 =  \rho \sigma\,,\quad \rho \equiv
	\frac{\beta}{2\pi}\,,\quad r=1\,.
\end{equation}
has an induced metric
\begin{equation}
    g_{\alpha\beta} = Q^2\ell^4 \rho^2 \eta_{\alpha\beta}\,
\end{equation}
and is geometrically trivial, \ie \eqref{Hage:trivial.geom} holds. There are again six massless and two massive scalars (associated with the cigar directions) with the mass parameters given by
\begin{equation}
    m^2 = \frac{6}{Q^2\ell^4}\,,\quad \tilde m^2 = 6\frac{\rho^2}{1+h_1}\,.
\end{equation}
From \eqref{Hage:Hagedorn.temp}, we can extract the dimensionless Hagedorn temperature,
\begin{equation}
\label{Ex:T11.rho_H}
	\rho_H = \frac1{\sqrt{\pi T_s}} \left[ 1 - \sqrt{\frac{3}{2}}\frac{1}{\sqrt{2\pi T_s}} 
	+ \frac{1+8\ln 2}{2} \left(\sqrt{\frac{3}{2}}\frac{1}{\sqrt{2\pi T_s}} \right)^2 +\cdots\right]~,\quad 2\pi\alpha^\prime\,T_s = Q^2\ell^4\,.
\end{equation}
It agrees with the expression \eqref{Ex:AdS.rho_H} for the $AdS_5\times S^5$ SUSY soliton, suggesting a sort of universality for the family of confining supergravity backgrounds presented in \cite{Chatzis:2024kdu,Chatzis:2024top}. It would be interesting to analyze this property further. Moreover, following a similar approach to the one in \cite{Castellani:2024ial,Castellani:2024pmx}, it would be also valuable asking how the thermal properties of this class of backgrounds change after (marginal) deformations of the solutions and how they depend on the dynamics of the Kaluza-Klein massive modes arising from the circle reductions.  We leave these topics for future studies. 

\section{Conclusions}
\label{concl}

In this paper, we have considered aspects of standard and non-standard approaches to the semiclassical quantization of the Green-Schwarz superstring. For the standard semiclassical approach, we have discussed the cancellation of the divergences in the  effective action for general (type-IIA) backgrounds, filling a gap in the literature. Our findings indicate that there is a preferred measure for the path integral over the fluctuations, and it would be interesting to investigate its implications on the (finite part of) the effective action for specific cases. 
Moreover, we have confirmed that the Nambu-Goto and Polyakov formulations are equivalent in any background, as far as the standard semiclassical quantization is concerned. 

Then, we have considered a non-standard semiclassical approach in the Polyakov formulation, in which the background auxiliary metric and the background induced metric are not conformal to each other, \ie the auxiliary metric is classically off-shell. Such an approach is motivated by recent developments in the holographic computation of the Hagedorn temperature in a large class of strongly-coupled confining gauge theories \cite{Bigazzi:2023oqm, Bigazzi:2024biz}, in which a classical scale parameter (the inverse temperature) is typically so small that it is comparable in size with the quantum fluctuations, making the standard approach unreliable. 
We have discussed the difficulties that one generally finds in such a setup. Amongst these, we have found a large degree of ambiguity for the semiclassical action, caused by the possibility of performing second-order field redefinitions of the auxiliary metric. We recall, however, that a first-order field redefinition is generally needed in order to absorb a first-order Lagrangian that remains, because the background auxiliary metric is off-shell. Field redefinitions and, consequently, ambiguities are avoided in such situations, in which the background worldsheet has vanishing extrinsic curvature. Fortunately, the setups relevant for the Hagedorn temperature we are interested in fall into this class of configurations.
Another difficulty arises from the fact that the fermionic kappa symmetry removes half of the fermions only on-shell, so that the reduction of the fermionic content of the fluctuations to eight 2-d spinors is not straightforward. We have proposed a resolution to this issue by first changing the systematics of the semiclassical expansion in the fermionic part of the action in such a way that the string vielbein is kept intact and is then formally replaced by a classical string vielbein that is consistent with the auxiliary metric. We remark that the same issue should be encountered whenever the auxiliary and intrinsic metrics are formally independent of each other, for example in conformal gauge, and we suspect that such a resolution has always been implicit in ``kappa symmetry gauge-fixed'' Lagrangians with an independent auxiliary metric, although we have not explicitly encountered it in the literature.

As an application of the non-standard semiclassical approach, we have presented a self-consistent construction of the vacuum configuration of a closed string that winds once around a flat compact direction in certain supergravity solutions. Such geometries can be viewed as setups with a finite temperature. We find that, generally, the vacuum string metric is not conformal to the induced metric of the classical string worldsheet, but they become essentially conformal to each other at very low temperatures. The temperature, at which one of the metrics (auxiliary or induced, depending on how the calculation is set up)
degenerates is the Hagedorn temperature. We have included a number of examples, to which this calculation applies. Our results reproduce the NLO term (in the expansion in terms of the inverse string tension) known from integrability-based calculations, and also provides an important input for the NNLO term in the effective approach. 
It would be interesting to ask under which conditions and how our calculation can be generalized to other background string worldsheets, although we suspect that the very notion of ``vacuum'' requires time translation invariance.

\section*{Acknowledgments}

We wish to thank Francesco Bigazzi, Aldo Cotrone,  Valentina Giangreco M. Puletti, Carlos N\'u\~nez, and Dima Sorokin for fruitful discussions. W.M.\ is grateful to the INFN Section of Firenze for kind hospitality. This work was partially supported by the INFN research initiatives GAST and STEFI. T.~C.~acknowledges support by the Simons Foundation grant 994300 (“Simons Collaboration on Confinement and QCD Strings”).

\begin{appendix}
\section{Geometry of embedded manifolds}
\label{embed}
\subsection{Structure Equations}
\label{embed:structure}
The differential geometry of embedded manifolds has been described in detail by Eisenhart \cite{Eisenhart}.  
We will present it here following \cite{Sorokin:1999jx} in a way involving forms, which naturally extends the embedding formalism to spinors and gives rise to the notion of Lorentz vector harmonics \cite{Galperin:1984av, Sokatchev:1985tc, Sokatchev:1987nk, Bandos:1990ji, Bandos:1990pk, Bandos:1995zw}.

Consider a $d$-dimensional (pseudo-)Riemannian manifold $\mathcal{M}_d$ embedded in a $D$-dimensional (pseudo-)Riemannian manifold $\mathcal{M}_D$. $\mathcal{M}_d$ and $\mathcal{M}_D$ will be called for short the \emph{worldsheet} and the \emph{spacetime}, respectively, with an eye to our application to strings. 
Our notation will be as follows. Greek letters will be used for curved (coordinate) indices, Latin letters for flat (frame) indices. Letters from the middle of the alphabet ($\mu,\nu,\ldots$ or $m,n,\ldots$) belong to spacetime, those from the beginning of the alphabet ($\alpha,\beta,\ldots$ or $a,b,\ldots$) belong to the worldsheet. In addition, letters $i,j,k$ belong to the normal bundle, for which there is no need for curved indices. Frame indices are raised and lowered with the (pseudo-)Euclidean metrics $\eta_{ab}$, $\eta_{mn}$, and $\eta_{ij}$, with the appropriate signatures. We will not be picky about the specific signatures and call the local rotation symmetries of the three frames $SO(d)$, $SO(D)$, and $SO(D-d)$, as appropriate.

So, let spacetime be parameterized by coordinates $x^\mu$ with metric $g_{\mu\nu}(x)$. Let us also introduce a local frame $E^m = \rmd x^\mu E_\mu{}^m$ such that
\begin{equation}
\label{embed:frame.props}
	E_\mu{}^m E_\nu{}^n  \eta_{mn} = g_{\mu\nu}~, \qquad g^{\mu\nu}  E_\mu{}^m E_\nu{}^n= \eta^{mn}~.
\end{equation}  
The second property defines the frame as orthonormal. 

Spacetime geometry is described by the torsion-free Cartan structure equations\footnote{It is not necessary to include torsion here, because the torsion structure is independent of the metric structure. If needed, torsion can be added to the spin connection by letting $\bar{\omega}_{ab} = \omega_{ab} + C_{ab}$, where $\omega_{ab}$ remains the unique torsion-free spin connection, and $C_{ab}=C_{[ab]}$ is the contorsion one-form. The first Cartan equation gives the torsion two-form, $T^a = e^b \wedge C_b{}^a$.}
\begin{align}
\label{embed:Cartan1}
	\rmd E^m + E^n \wedge \Omega_n{}^m &=0~,\\
\label{embed:Cartan2}
	\rmd \Omega_m{}^n + \Omega_m{}^p \wedge \Omega_p{}^n &=R_m{}^n~,
\end{align}
with $\Omega_m{}^n$ and $R_m{}^n$ being the connection one-forms and curvature two-forms, respectively.
They satisfy the Bianchi identities\footnote{We use the convention that $\rmd$ acts from the right.}
\begin{align}
\label{embed:Bianchi1}
	E^n \wedge R_n{}^m&=0~,\\
\label{embed:Bianchi2}
	\rmd R_m{}^n + R_m{}^p \wedge \Omega_p{}^n - \Omega_m{}^p \wedge R_p{}^n &=0~.
\end{align}

The embedding of the worldsheet in spacetime is defined by functions $x^\mu(\xi)$, where $\xi^\alpha$ are coordinates on the worldsheet. Thus, the derivatives 
\begin{equation}
\label{embed:tangents}
  x^\mu_\alpha(\xi) \equiv \partial_\alpha x^\mu(\xi)
\end{equation}
are the tangents that provide the pull-backs of spacetime tensors onto the worldsheet. In particular, the induced metric is 
\begin{equation}
\label{embed:g}
  g_{\alpha\beta} = x^\mu_\alpha x^\nu_\beta\, g_{\mu\nu}~. 
\end{equation}

The tangents $x_\alpha^\mu$ do not suffice to define a worldsheet local frame $e^a$. If we introduce the pull-backs of the spacetime frame one-forms onto the worldsheet,
$$
	\pull{E^m} \equiv \rmd \xi^\alpha x_\alpha^\mu E_\mu{}^m~,
$$
of which there are a total of $D$, we find that only $d$ of these are linearly independent. There always exists a local $SO(D)$ rotation $u_m{}^n(\xi)$, with the properties
\begin{equation}
\label{embed:u.props}
  u_m{}^p u_n{}^q \eta_{pq} = \eta_{mn}~, \qquad \eta^{mn} u_m{}^p u_n{}^q = \eta^{pq}~, 
\end{equation}
such that the rotated frame ${E'}^n = E^m u_m{}^n$, which we may call a worldsheet adapted frame, contains $D-d$ basis one-forms whose pull-backs vanish. The remaining $d$ basis one-forms can be identified as the worldsheet frame. Writing $u_m{}^n = (u_m{}^a, u_m{}^i)$, this means 
\begin{equation}
\label{embed:frame.ws}
	\pull{E^m} u_m{}^a = e^a~,\qquad  \pull{E^m} u_m{}^i = e^i = 0~. 
\end{equation}
The $u_m{}^n$ with these properties are called the Lorentz vector harmonics.
The quantities
\begin{equation}
\label{embed:N.def} 
	N_\mu^i \equiv E_\mu{}^m u_m{}^i 
\end{equation}
are the normals to the worldsheet, satisfying
\begin{equation}
\label{embed:N.props} 
	x_\alpha^\mu N_\mu^i =0~,\qquad g^{\mu\nu} N_\mu^iN_\nu^j = \eta^{ij}~. 
\end{equation}

It is easy check that the frame $e^a$ reproduces the induced metric \eqref{embed:g}, 
$$ 
	\pull{g} = \pull{E^m} \otimes \pull{E^n} \eta_{mn} 
	= \pull{E^m} \otimes \pull{E^n} u_m{}^p u_n{}^q \eta_{pq} 
	= \pull{E^m} \otimes \pull{E^n} u_m{}^a u_n{}^b \eta_{ab} = e^a \otimes e^b \eta_{ab}~.
$$
In the third equality we have used $\eta_{mn}=\mathrm{diag}(\eta_{ab}, \eta_{ij})$ and then \eqref{embed:frame.ws}. This trick will be useful frequently. 
The proof that the $e^a$ form an orthonormal basis with respect to the induced metric, $g^{\alpha\beta} e_\alpha{}^a e_\beta{}^b=\eta^{ab}$, proceeds as usual.

The properties \eqref{embed:u.props} imply that the inverse $SO(D)$ matrix $(u^{-1})_m{}^n$ is given by
\begin{equation}
\label{embed:u.inverse}
	(u^{-1})_m{}^n = \eta^{np} u_p{}^q \eta_{qm} = u^n{}_m~.
\end{equation}
Therefore, \eqref{embed:frame.ws} yields more explicitly
\begin{equation}
\label{embed:Em}
	\pull{E^m} = \pull{E^n} u_n{}^q u^m{}_q = e^a u^m{}_a~.
\end{equation}  

Let us analyze the geometry of the embedding, which is described by Cartan's structure equations on the worldsheet. We start with 
\begin{equation}
\label{embed:Cartan1.ws}
	\rmd e^a + e^b\wedge\omega_b{}^a = 0~.
\end{equation}
Using \eqref{embed:Em}, \eqref{embed:Cartan1.ws} leads to the relation 
\begin{equation}
\label{embed:Cartan1.ws.cond}
	e^b u^m{}_b \wedge \left( \rmd u_m{}^a -\pull{\Omega_m{}^n} u_n{}^a + u_m{}^c \omega_c{}^a \right) =0~,
\end{equation} 
which implies the equation of Gauss\footnote{Another term of the form $u_m{}^b \Lambda_b{}^a$ on the right hand side is not allowed by symmetry. To satisfy \eqref{embed:Cartan1.ws.cond}, one would need $\Lambda_b{}^a = e^c \Lambda_{cb}{}^a$ with $\Lambda_{cb}{}^a = \Lambda_{bc}{}^a$, but also $\Lambda_{cba} = - \Lambda_{cab}$ by the symmetry properties of the spin connection. Hence, $\Lambda_{cba} = \Lambda_{bca} = - \Lambda_{bac} = - \Lambda_{abc} = \Lambda_{acb} = \Lambda_{cab} = -\Lambda_{cba}$, so $\Lambda_{abc}=0$.} 
\begin{equation}
\label{embed:Gauss1.ws}
	 \rmd u_m{}^a -\pull{\Omega_m{}^n} u_n{}^a + u_m{}^b \omega_b{}^a = u_m{}^i \tilde{K}_{ib} \eta^{ba}~, 
\end{equation}
where $\tilde{K}_{ib}$ are one-forms.

Contracting \eqref{embed:Gauss1.ws} with $u^m{}_b$ determines the worldsheet spin connections,
\begin{equation}
\label{embed:spin.conn.ws}
	\omega_a{}^b = -u^m{}_a \rmd u_m{}^b 
	+ u^m{}_a \pull{\Omega_m{}^{n}} u_n{}^b~.
\end{equation} 
Instead, contracting \eqref{embed:Gauss1.ws} with $u^m{}_i$ yields the one-forms $\tilde{K}_{ia}$, 
\begin{equation}
\label{embed:K.i}
	\tilde{K}_{i}{}^a = u^m{}_i \rmd u_m{}^a - u^m{}_i\pull{\Omega_m{}^n} u_n{}^a ~.
\end{equation}

Similarly, let us consider the exterior derivative of the second equation of \eqref{embed:frame.ws}. It gives rise to the condition 
$$ 
	e^a u^m{}_a \wedge \left( \rmd u_m{}^i - \pull{\Omega_m{}^n} u_n{}^i \right)=0~,
$$
which is solved by 
\begin{equation}
\label{embed:Weingarten.ws}
	 \rmd u_m{}^i -\pull{\Omega_m{}^n} u_n{}^i + u_m{}^j A_j{}^i  = -u_m{}^a K_a{}^i~,
\end{equation}
with some one-forms $A_j{}^i$ and $K_a{}^i$, where the latter must have symmetric coefficients, 
\begin{equation}
\label{embed:K.symm}
	K_a{}^i= e^b K_{ba}{}^i~,\qquad K_{ab}{}^i = K_{(ab)}{}^i~.
\end{equation}
Contracting \eqref{embed:Weingarten.ws} with $u^m{}_a$ and $u^m{}_i$ yields, respectively, 
\begin{equation}
\label{embed:K.A.rels}
	\tilde{K}_{ia} = K_{ai}~, \qquad A^{(ij)}=0~.
\end{equation}
$K_{ai}$ is the second fundamental form, which measures the extrinsic curvature of the worldsheet embedding. $A_i{}^j$ is the normal bundle connection. The left hand sides of \eqref{embed:Gauss1.ws} and \eqref{embed:Weingarten.ws} suggest to define the worldsheet covariant differential $\nabla$ such that these equations simply become 
\begin{equation}
\label{embed:cov.nabla}
	\nabla u_m{}^a = u_m{}^i K^a{}_i~,\qquad \nabla u_m{}^i = - u_m{}^a K_a{}^i~.
\end{equation}

Finally, the integrability conditions of \eqref{embed:Gauss1.ws} and \eqref{embed:Weingarten.ws} are the equations of Gauss, Codazzi and Ricci. These are, respectively,
\begin{align}
\label{embed:gauss2} 
	u^m{}_a R_m{}^n u_n{}^b &= R_a{}^b - K_{ai} \wedge K^{bi}~,\\
\label{embed:codazzi}
	u^m{}_i R_m{}^n u_n{}^a &= -\left( \rmd K^a{}_i + K^b{}_i \wedge \omega_b{}^a + A_i{}^j \wedge K^a{}_j\right)~, \\
\label{embed:ricci}
	u^m{}_i R_m{}^n u_n{}^j &= \rmd A_i{}^j + A_i{}^k \wedge A_k{}^j - K_{ai} \wedge K^{aj}~.
\end{align}

\subsection{Spinors}
\label{embed:spinors}
Here, we will discuss consider spinors on embedded manifolds. They deserve a more detailed discussion, because of the existence of different notions of spin structure on embedded manifolds and subtleties in the calculation of functional determinants that ensue from them \cite{Kavalov:1986nx, Sedrakian:1986np, Langouche:1987my, Langouche:1987mw, Langouche:1987mx, Wiegmann:1989md, Karakhanian:1990nd}, see also \cite{Drukker:2000ep}. The notion of Lorentz harmonics introduced above makes the discussion straightforward.

To start, consider the gamma-matrix valued one-forms $E^m \Gamma_m$. Their pull-backs onto the worldsheet, 
\begin{equation}
\label{embed:gamma.pullback}
	\pull{E^m}\Gamma_m = e^a u^m{}_a \Gamma_m~,
\end{equation}
would suggest to take the matrices $u^m{}_a \Gamma_m$ as worldsheet gamma matrices, but this is not correct. Although they satisfy the Clifford algebra on the worldsheet, they are generally not constant. Because $u^m{}_n$ is an $SO(D)$ transformation, there exists by definition an associated transformation $U$ in spinor space such that
\begin{equation}
\label{embed:U.def}
	u^m{}_n U\Gamma_m U^{-1} = \Gamma_n~, 
\end{equation}
or, equivalently,
\begin{equation}
\label{embed:U.def.2}
	U\Gamma_m U^{-1} = u_m{}^n \Gamma_n~. 
\end{equation}
For consistency, \eqref{embed:U.def} implies 
\begin{equation}
\label{embed:dU.eq}
	u^m{}_n (\rmd u_m{}^p) \Gamma_p = \left[ (\rmd U)U^{-1}, \Gamma_n\right]~,
\end{equation}
which is explicitly solved by
\begin{equation}
\label{embed:dU.sol}
	(dU)U^{-1} = \frac14 u^m{}_p (\rmd u_m{}^q) \Gamma_q{}^p~.
\end{equation}

Consider now the covariant derivative of a spacetime spinor $\theta$, 
\begin{eqnarray}
\label{embed:spin.cov.der}
	\rmD \theta = \left( \rmd + \frac14 \Omega^{mn} \Gamma_{nm} \right) \theta~.
\end{eqnarray}
Writing $\theta = U^{-1} U \theta$ and using \eqref{embed:U.def.2} and \eqref{embed:dU.sol}, one easily finds
\begin{equation}
\label{embed:spin.der.trafo}
	\rmD\theta = U^{-1} \left[ \rmd -\frac14 u^{pn} \left( \rmd u_p{}^m - \Omega_p{}^q u_q{}^m\right) \Gamma_{mn} \right] U\theta~.
\end{equation}
Of course, \eqref{embed:spin.der.trafo} gives the transformation rule of the spin connections under local frame rotations, but in our context it also provides the pull-back of $\rmD \theta$ onto the worldsheet. Using \eqref{embed:Gauss1.ws} and \eqref{embed:Weingarten.ws}, one obtains
\begin{equation}
\label{embed:spin.der.pullback}
	\pull{\rmD\theta} = U^{-1} \left( \pull{\rmd} -\frac14 \omega^{ab} \Gamma_{ab} -\frac14 A^{ij} \Gamma_{ij} -\frac12 K^{ai} \Gamma_{ai} \right) U\theta~.
\end{equation} 
One may rewrite this concisely as 
\begin{equation}
\label{embed:spin.der.pullback.cov}
	\pull{\rmD\theta} = U^{-1} \left( \nabla -\frac12 K^{ai} \Gamma_{ai} \right) U\theta~,
\end{equation} 
where we have defined the worldsheet covariant derivative $\nabla$ including the internal and normal bundle connections. It satisfies
\begin{equation}
\label{embed:gamma.cov.der}
	\nabla \Gamma_a = 0~, \qquad \nabla \Gamma_i =0~,
\end{equation}
where it is intended, as usual, to act on the indices and on both the left and right of the matrices in spinor space. So, we see that $\psi = U \theta$ is the spinor the induced worldsheet covariant derivative acts on.

 \section{Type IIA supergravity field equations}
\label{sugra}

The bosonic sector of type IIA supergravity consists of the metric $g_{\mu\nu}$, the dilaton $\phi$, the Kalb-Ramond field $B_{\mu\nu}$ with its associated field strength $H_3 = \rmd B_2$, as well as the Ramond-Ramond potentials $C_1$ and $C_3$. Their associated field strengths are, including the Chern-Simons terms, 
\begin{equation}
\label{sugra:F.RR}
	F_2 = \rmd C_1~,\qquad F_4 = \rmd C_3 - C_1 \wedge H_3~.
\end{equation}
The field equation for the metric is the Einstein equation 
\begin{equation}
\label{sugra:einstein}
	R_{\mu\nu} - \frac14 H_{\mu\rho\lambda} H_\nu{}^{\rho\lambda} -\frac12 \e{2\phi} \left[ \left| F_4\right|^2_{\mu\nu} +\left| F_2\right|^2_{\mu\nu} -\frac12 g_{\mu\nu} \left(\left| F_4\right|^2 + \left| F_2\right|^2 \right) \right] = -2\nabla_\mu \nabla_\nu \phi~, 
\end{equation}
where we have abbreviated 
\begin{equation}
\label{sugra:F2.abbrev}
	\left| F_n\right|^2_{\mu\nu} = \frac1{(n-1)!} F_{\mu\rho_1\ldots \rho_{n-1}}F_\nu {}^{\rho_1\ldots \rho_{n-1}}~,\qquad
	\left| F_n\right|^2 = \frac1{n!} F_{\rho_1\ldots \rho_n}F^{\rho_1\ldots \rho_n}~.
\end{equation}
Then, there are the Maxwell equations
\begin{align}
\label{sugra:Maxwell.F2}
	\rmd \ast F_2 &= H_3 \wedge \ast F_4~,\\
\label{sugra:Maxwell.F4}
	\rmd \ast F_4 &= - H_3 \wedge F_4~,\\
\label{sugra:Maxwell.H3}
	\rmd \left(\e{-2\phi} \ast H_3 \right) &= \frac12 F_4 \wedge F_4 - F_2 \wedge \ast F_4~,
\end{align}
as well as the dilaton field equation
\begin{equation}
\label{sugra:dilaton}
	\nabla^\mu \nabla_\mu \phi - 2 (\nabla^\mu \phi)(\nabla_\mu \phi) = -\frac12 \left|H_3\right|^2 + \frac14 \e{2\phi} 
	\left( \left|F_4\right|^2 +3 \left|F_2\right|^2 \right)~.
\end{equation}
Because we will need it in the main text, we provide here \eqref{sugra:Maxwell.H3} in coordinate form,
\begin{equation}
\label{sugra:Maxwell.h3.coord}
	-\nabla^\lambda \left( \e{-2\phi} H_{\lambda\mu\nu} \right) = 
	\frac1{2(4!)^2} \epsilon_{\mu\nu\rho_1\cdots \rho_8} F^{\rho_1\cdots \rho_4} F^{\rho_5\cdots \rho_8} + \frac12 F_{\mu\nu\rho\lambda} F^{\rho\lambda}~.
\end{equation}

 \section{Canonical quantization}
\label{quant}

In this appendix, we perform the canonical quantization of 2-d scalars and fermions on a worldsheet with the topology of a  cylinder, $\tau\in (-\infty,\infty)$, $\sigma\in [0,2\pi)$, constant (auxiliary) metric and constant mass parameters.
To keep the discussion as general as possible, this will be done with a generic inverse metric $\unih^{\alpha\beta}$ parameterized by two constant coefficients $h_1$ and $h_2$ as follows,
\begin{equation} 
\label{quant:h.param}
	\unih^{\alpha\beta} = \frac1{H} \begin{pmatrix}
	-1+h_1 & h_2 \\
	h_2 & 1+h_1 
	\end{pmatrix}~, \qquad H = \sqrt{1-h_1^2+h_2^2}~.
\end{equation}
It is assumed that the argument under the square root in $H$ is positive. The metric $\unih_{\alpha\beta}$
is 
\begin{equation} 
\label{quant:h.down}
	\unih_{\alpha\beta} = \frac1{H} \begin{pmatrix}
	-1-h_1 & h_2 \\
	h_2 & 1-h_1 
	\end{pmatrix}~.
\end{equation}
An inverse zweibein compatible with \eqref{quant:h.param} is given by
\begin{equation}
\label{quant:zweibein.convenient}
	\unie^\alpha{}_a = \frac1{\sqrt{2H(H+1)}} 
	\begin{pmatrix} H+1-h_1& -h_2 \\ h_2 & H+1+h_1 \end{pmatrix}~.
\end{equation}

\subsection{Massless scalars}
\label{scal:massless}

The dynamics of a massless scalar is governed by the action 
\begin{equation}
\label{quant:S.massless}
	S = -\frac1{4\pi\alpha'} \int\rmd^2 \sigma\, \unih^{\alpha\beta} (\partial_\alpha \chi) (\partial_\beta \chi)~.
\end{equation}
With the parameterization of $\unih^{\alpha\beta}$ given in \eqref{quant:h.param}, this gives rise to the field equation
\begin{equation}
\label{quant:massless.eom}
	\left[ (-1+h_1) \partial_\tau^2 +2h_2\partial_\tau\partial_\sigma + (1+h_1) \partial_\sigma^2 \right] \chi =0~. 
\end{equation}
For periodic boundary conditions on $\sigma$, the solution can be expressed as 
\begin{equation}
\label{quant:massless.mode.expansion}
	\chi(\tau,\sigma) = \chi_0 + \alpha'\frac{H}{1-h_1} q_0 \tau 
	+ \sqrt{\frac{\alpha'}{2}} \sum\limits_{n\neq 0} \frac{i}{n} \left[ \chi_n \e{-in(\nu_+\tau +\sigma)} 
	+ \tilde{\chi}_n \e{-in(\nu_-\tau -\sigma)}\right]~, 
\end{equation}
where $n\in \mathbb{Z}$,
\begin{equation}
\label{quant:nu.pm}
	\nu_{\pm} = \frac{H\pm h_2}{1-h_1}~,
\end{equation}
and we have normalized the mode coefficients such that they satisfy
\begin{equation}
\label{quant:massless.commutators}
	[\chi_0,q_0] = i~,\qquad [ \chi_n, \chi_m] = [\tilde{\chi}_n,\tilde{\chi}_m] = n \delta_{n,-m}
\end{equation}
upon canonical quantization. The canonical momentum is 
\begin{equation}
\label{quant:momentum}
	\pi(\tau,\sigma) = \frac{\delta S}{\delta (\partial_\tau \chi)} = -\frac{1}{2\pi\alpha'} \left( \unih^{\tau\tau}\partial_\tau \chi + \unih^{\tau\sigma} \partial_\sigma \chi \right) 
	= \frac{(1-h_1) \partial_\tau \chi -h_2\partial_\sigma \chi}{2\pi\alpha'H}~,
\end{equation}
which results in
\begin{equation}
\label{quant:massless.momentum}
	\pi(\tau,\sigma)= \frac1{2\pi} q_0 
	+ \frac1{2\pi\sqrt{2\alpha'}} \sum\limits_{n\neq 0} \left[ \chi_n \e{-in(\nu_+\tau +\sigma)} 
	+ \tilde{\chi}_n \e{-in(\nu_-\tau -\sigma)}\right]~. 
\end{equation}
Reality of $\chi$ demands $\chi_n^\dagger = \chi_{-n}$ and, as usual, the mode coefficients with $n$ positive annihilate the vacuum.

The contribution of a massless scalar to the tensor $t_{\alpha\beta}$ in \eqref{Hage:t.def} is found to be 
\begin{equation}
\label{quant:massless.tab}
	t_{\alpha\beta} = \vev{(\partial_\alpha \chi)(\partial_\beta \chi)} = 
	-\frac{\alpha'}{12}  T_{\alpha\beta}~.
\end{equation}
To obtain this result, one must normal order the mode coefficients and sum up the normal ordering constant using $\sum_{n=1}^\infty n =\zeta(-1) = -\frac1{12}$. The tensor $T_{\alpha\beta}$ is  
\begin{equation}
\label{quant:T.ab.def}
	T_{\alpha\beta} = \begin{pmatrix} \frac{H^2+h_2^2}{(1-h_1)^2} & \frac{h_2}{1-h_1} \\ \frac{h_2}{1-h_1} & 1 \end{pmatrix}~,
\end{equation}
which is traceless, $\unih^{\alpha\beta} T_{\alpha\beta} =0$.

\subsection{Massive scalars}
\label{scal:massive}

The dynamics of a massive scalar is governed by the action \eqref{Hage:action.scalars}
\begin{equation}
\label{quant:S.massive}
	S = -\frac1{4\pi\alpha'} \int\rmd^2 \sigma \left[ \unih^{\alpha\beta} (\partial_\alpha \chi) (\partial_\beta \chi) 
	+  \mu_b^2 \chi^2 \right]~.
\end{equation}
With $\unih^{\alpha\beta}$ given by \eqref{quant:h.param}, this gives rise to the field equation
\begin{equation}
\label{quant:massive.eom}
	\left[ (-1+h_1) \partial_\tau^2 +2h_2\partial_\tau\partial_\sigma + (1+h_1) \partial_\sigma^2 - H \mu_b^2 \right] \chi =0~. 
\end{equation}
The mode expansion of the solution is 
\begin{equation}
\label{quant:massive.mode.expansion}
	\chi(\tau,\sigma) = i \sqrt{\frac{\alpha'}{2}} \sum\limits_{n=-\infty}^\infty \frac{1}{\sqrt{\lambda_n}} 
	\left[ a_n \e{-i(\omega_n \tau + n \sigma)} - a_n^\dagger \e{i(\omega_n \tau + n \sigma)}\right]~, 
\end{equation}
with $n\in \mathbb{Z}$, the mode frequencies given by
\begin{equation}
\label{quant:omega.n}
	\omega_n = \frac{H}{1-h_1} \lambda_n +\frac{n h_2}{1-h_1}~,
\end{equation}
and 
\begin{equation}
\label{quant:lambda.n}
	\lambda_n = \sqrt{n^2 + \tilde{m}_b^2}~, \qquad \tilde{m}_b^2 = \frac{1-h_1}{H} \mu_b^2.
\end{equation}

The mode coefficients are the usual ladder operators satisfying 
\begin{equation}
\label{quant:massive.commutators}
	[a_m,a_n^\dagger] = \delta_{m,n}~.
\end{equation}
For completeness, from \eqref{quant:momentum} we get the canonical momentum 
\begin{equation}
\label{quant:massive.momentum}
	\pi(\tau,\sigma) = \frac1{2\pi\sqrt{2\alpha'}} \sum\limits_{n=-\infty}^\infty \sqrt{\lambda_n}
	\left[ a_n \e{-i(\omega_n \tau + n \sigma)} + a_n^\dagger \e{i(\omega_n \tau + n \sigma)}\right]~. 
\end{equation}

The contribution of a massive scalar to the tensor $t_{\alpha\beta}$ in \eqref{Hage:t.def} is 
\begin{equation}
\label{quant:t.massive.scalar}
	t_{\alpha\beta} = \vev{(\partial_\alpha \chi)(\partial_\beta\chi) + \frac12 \mu_b^2 \unih_{\alpha\beta} \chi^2}  +
	\frac{\mu_b^2}{2\Omega} \left( \unig_{\alpha\beta} - \Omega \unih_{\alpha\beta} \right) \vev{\chi^2}~,
\end{equation}
where the second term is manifestly traceless (under contraction with $\unih^{\alpha\beta}$). 
The first term yields
\begin{equation}
\label{quant:massive.tab1}
	\vev{(\partial_\alpha \chi)(\partial_\beta\chi) + \frac12 \mu_b^2 \unih_{\alpha\beta} \chi^2} = 
	-\alpha' \left(\Delta_b -\tilde{m}_b^2 \Delta_b' \right) T_{\alpha\beta}~, 
\end{equation}
where $T_{\alpha\beta}$ is again given by \eqref{quant:T.ab.def}, and the functions $\Delta_b(\tilde{m}_b)$ and $\Delta_b'(\tilde{m}_b)$ are the infinite sums 
\begin{align}
	\label{quant:delta.b}
	\Delta_b &=-\frac12 \sum\limits_{n=-\infty}^\infty \lambda_n 
	= -\frac{\tilde{m}_b}2 - \sum\limits_{n=1}^\infty \sqrt{n^2+\tilde{m}_b^2}~,\\
	\label{quant:delta.prime.b}
	\Delta'_b &=-\frac14 \sum\limits_{n=-\infty}^\infty \frac{1}{\lambda_n} 
	= \frac{\rmd \Delta_b}{\rmd \tilde{m}_b^2}~.
\end{align}
The expectation value in the second term in \eqref{quant:t.massive.scalar} is
\begin{equation}
\label{quant:scal.vev.chi2}
	\vev{\chi^2} = -2\alpha' \Delta'_b~, 
\end{equation}
hence
\begin{equation}
\label{quant:t.scalar.result}
	t_{\alpha\beta} =-\alpha' \left[ \left(\Delta_b -\tilde{m}_b^2 \Delta_b' \right) T_{\alpha\beta} 
	+\mu_b^2 \Delta_b' \left( \Omega^{-1} \unig_{\alpha\beta} - \unih_{\alpha\beta}\right) \right]~. 
\end{equation} 
It is evident that the $t_{\alpha\beta}$ is traceless, $\unih^{\alpha\beta} t_{\alpha\beta}=0$.  
The result \eqref{quant:massless.tab} for a massless scalar is reproduced by setting $\mu_b=0$ and noting that $\Delta_b(0)=-\zeta(-1)=\frac1{12}$.

\subsection{Fermions}
\label{quant:fermions}

The action \eqref{Hage:action.fermions} for a fermion is
\begin{equation}
\label{quant:S.fermion}
	S = -\frac{i}{2\pi\alpha'} \int\rmd^2 \sigma \bar{\psi} \left( \unie^\alpha{}_a \gamma^a \partial_\alpha + \mu_f \right) \psi~.
\end{equation}

To be explicit, we use the representation of the (frame) gamma matrices $\gamma^a$
\begin{equation}
\label{quant:ferm.gamma.mat}
	\gamma^0 = \begin{pmatrix} 0& 1 \\ -1 & 0 \end{pmatrix}~,\qquad
	\gamma^1 = \begin{pmatrix} 0& 1 \\ 1 & 0 \end{pmatrix}~,
\end{equation}
and the zweibeins \eqref{quant:zweibein.convenient}. The fermions satisfy antiperiodic boundary conditions on the compact cycle. 
The corresponding mode expansion can be written in the form
\begin{equation}
\label{quant:psi.sol}
	\psi = \sum\limits_{r} \left[ c_r u_r \e{-i(\omega_{r+}\tau +r\sigma)} + d_r v_r  \e{-i(\omega_{r-}\tau -r\sigma)} \right]~,
\end{equation} 
with $r\in \mathbb{Z}+\frac12$, the mode frequencies  
\begin{equation}
\label{quant:omega.r}
	\omega_{r\pm} = \frac{H}{1-h_1} \sgn r\, \lambda_r \pm \frac{h_2r}{1-h_1}~,
\end{equation}
where
\begin{equation}
\label{quant:lambda.r}
	\lambda_r = \sqrt{r^2 +\tilde{m}_f^2}~, \qquad \tilde{m}_f^2 = \frac{1-h_1}{H} \mu_f^2~.
\end{equation}
It is evident that these are the analogues of \eqref{quant:omega.n} and \eqref{quant:lambda.n}.

The mode spinors in \eqref{quant:psi.sol} are given by 
\begin{align}
\label{quant:u}
	u_r &= \frac{\sqrt{\alpha'}}{\sqrt{Z(H+1-h_1+h_2)}} 
	\begin{pmatrix}
	i (H+1-h_1+h_2) \omega_{r+} + i(H+1+h_1-h_2)r \\
	\sqrt{2H(H+1)} \mu_f
	\end{pmatrix}~, \\
	v_r &= \frac{\sqrt{\alpha'}}{\sqrt{Z(H+1-h_1-h_2)}}
	\begin{pmatrix}
	-\sqrt{2H(H+1)} \mu_f \\
	i (H+1-h_1-h_2) \omega_{r-} + i(H+1+h_1+h_2)r 
	\end{pmatrix}~,
\end{align}
with the normalization constant 
\begin{equation}
\label{quant:Znorm}
	Z = \frac{4 H\sqrt{2H(H+1)} (\lambda_r + |r|) \lambda_r}{1-h_1}~.
\end{equation}
The Majorana condition implies that the mode operators satisfy $c_{-r}= c_r^\dagger$, $d_{-r}= d_r^\dagger$. They also satisfy the canonical anticommutation relations
\begin{equation}
\label{quant:anti.comm}
	\left\{ c_r, c_{r'}\right\} = \delta_{r,-r'}~, \qquad 
	\left\{ d_r, d_{r'}\right\} = \delta_{r,-r'}~.
\end{equation}

The contribution of a fermion to the tensor $t_{\alpha\beta}$ in \eqref{Hage:t.def} is 
\begin{equation}
\label{quant:t.fermion}
	t_{\alpha\beta} = i\vev{\bar{\psi} \left( \gamma^a \unie_{a(\alpha} \partial_{\beta)} +\frac12 \mu_f \unih_{\alpha\beta}\right) \psi} + i\frac12 \mu_f\left( \Omega^{-1} \unig_{\alpha\beta} -\unih_{\alpha\beta}\right) \vev{ \bar{\psi} \psi}~,
\end{equation}
where the second term is again manifestly traceless. 
The first term in yields
\begin{equation}
\label{quant:fermion.tab1}
	i\vev{\bar{\psi} \left( \gamma^a \unie_{a(\alpha} \partial_{\beta)} +\frac12 \mu_f \unih_{\alpha\beta}\right) \psi} 
	= \alpha' \left(\Delta_f -\tilde{m}_f^2 \Delta_f' \right) T_{\alpha\beta}~, 
\end{equation}
where $T_{\alpha\beta}$ is again given by \eqref{quant:T.ab.def}, and the functions $\Delta_f$ and $\Delta_f'$ are the infinite sums 
\begin{align}
	\label{quant:delta.f}
	\Delta_f(\tilde{m}_f) &= -\sum\limits_{r>0} \lambda_r= -\sum\limits_{n=0}^\infty \sqrt{\left(n+\frac12\right)^2+\tilde{m}_f^2}~,\\
	\label{quant:delta.prime.f}
	\Delta'_f(\tilde{m}_f) &=-\frac12 \sum\limits_{r>0} \frac{1}{\lambda_r} 
	= \frac{\rmd \Delta_f(\tilde{m}_f)}{\rmd \tilde{m}_f^2}~.
\end{align}
The vacuum expectation value in the second term in \eqref{quant:t.fermion} is
\begin{equation}
\label{quant:ferm.vev.psi2}
	i \vev{ \bar{\psi} \psi} = 2\alpha' \mu_f \Delta'_f~,
\end{equation}
so that
\begin{equation}
\label{quant:t.fermion.result}
	t_{\alpha\beta} =\alpha' \left[ \left(\Delta_f -\tilde{m}_f^2 \Delta_f' \right) T_{\alpha\beta} 
	+\mu_f^2 \Delta_f' \left( \Omega^{-1} \unig_{\alpha\beta} - \unih_{\alpha\beta}\right) \right]~. 
\end{equation}

 \end{appendix}

\bibliographystyle{utphys}
\providecommand{\href}[2]{#2}\begingroup\raggedright\endgroup

\end{document}